\documentclass[aps,prl,reprint,nofootinbib,floatfix]{revtex4-2}
\usepackage{mathrsfs}
\usepackage{graphicx}
\usepackage{amsmath}
\usepackage{amsfonts}
\usepackage{amssymb}
\usepackage{color}

\usepackage{epsfig}
\usepackage{CJK}
\usepackage{graphicx}
\usepackage{epsfig}
\usepackage{eepic}
\usepackage{bbm}
\usepackage{dcolumn}
\usepackage{bm}
\usepackage[normalem]{ulem}

\usepackage{multirow}
\usepackage{slashed}

\newcommand{\omits}[1]{}

\def\bc{\begin{center}}

\def\ec{\end{center}}
\def\be{\begin{eqnarray}}
\def\ee{\end{eqnarray}}

\definecolor{dyellow}{rgb}{1.,0.8,.0}
\definecolor{myblue}{rgb}{.1,.1,.7}
\definecolor{dcyan}{rgb}{.0,.6,.6}
\definecolor{cyan}{rgb}{0.4,1.0,1.0}
\definecolor{dmagenta}{rgb}{0.6,0.0,0.6}
\definecolor{brown}{rgb}{0.6,0.2,0.}
\definecolor{darkblue}{rgb}{.0,.0,0.5}
\definecolor{darkred}{rgb}{0.75,0.0,0.0}
\definecolor{orange}{rgb}{1.,.6,.0}
\definecolor{dorange}{rgb}{0.8,.4,.0}
\definecolor{green}{rgb}{0.0,1.0,0.0}
\definecolor{darkgreen}{rgb}{0.0,0.6,0.0}
\definecolor{purple}{rgb}{.4,.0,.4}
\definecolor{lightgrey}{rgb}{0.7, 0.7, 0.7}
\definecolor{grey}{rgb}{0.4, 0.4, 0.4}

\newcommand{\nc}{\newcommand}
\nc{\rnc}{\renewcommand} \nc{\ket}[1]{\left | \, #1 \right \rangle}
\nc{\bra}[1]{\left \langle #1 \, \right |}
\nc{\ua}{\uparrow} \nc{\da}{\downarrow}

\nc{\braket}[2]{\langle\, #1\,|\,#2\,\rangle}
\nc{\half}{\frac{1}{2}}

\nc{\prj}{\mathcal{P}} \nc{\hilb}{\mathcal{H}}
\nc{\pth}{\mathcal{C}} \nc{\inprod}[2]{\braket{#1}{#2}}
\nc{\upket}{\ket{\uparrow}} \nc{\downket}{\ket{\downarrow}}
\nc{\upbra}{\bra{\uparrow}} \nc{\downbra}{\bra{\downarrow}}

\begin{document}


\title{Holographic Bit Threads from String-Diagrammatic Quantum Information Flow}

\author{Yi-Yu Lin$^{1,2}$} \email{yiyu@simis.cn}
\author{Song Cheng$^3$} \email{}

\affiliation{${}^1$Fudan Center for Mathematics and Interdisciplinary Study, Fudan University, Shanghai, 200433, China}
\affiliation{${}^2$Shanghai Institute for Mathematics and Interdisciplinary Sciences (SIMIS), Shanghai, 200433, China}
\affiliation{${}^3$Beijing Institute of Mathematical Sciences and Applications (BIMSA),
	Beijing, 101408, China}


\begin{abstract}

Bit threads, arising as the convex dual of the minimal-surface formula for holographic entanglement entropy, are line-like structures with nontrivial bulk trajectories. Their physical picture has long been associated with Bell-pair interpretations, yet the meaning of their detailed trajectories, particularly their nonuniqueness, remains unclear. We propose to apply Coecke’s notion of quantum information flow (QIF) in protocol-network geometry, developed within categorical quantum mechanics, to a holographic setup, and show that its trajectories within the discrete holographic bulk---the tensor network---obey the divergence-free and density-bound conditions of bit threads. From this process-centered perspective, bit-thread nonuniqueness becomes natural: for a given resource state, different protocols realizing the same  entanglement-distillation or quantum-message-transfer task can give rise to different QIF trajectories. We further analyze stabilizer-type holographic tensor networks using ZX string diagrams and show that QIF trajectories can probe finer entanglement structure beyond the level captured by entanglement entropy.

\end{abstract}


\maketitle


In holographic duality~\cite{Maldacena:1997re,Gubser:1998bc,Witten:1998qj}, spacetime geometry is related to quantum entanglement in a remarkably concrete way. The Ryu-Takayanagi formula~\cite{Ryu:2006bv,Ryu:2006ef,Hubeny:2007xt} states that, at leading order, the entanglement entropy of a boundary region $A$ is given by the area of the bulk minimal surface $\gamma_A$ homologous to $A$, $S(A)=\frac{\operatorname{Area}(\gamma_A)}{4G_N}.$
Using the max flow-min cut theorem, Freedman and Headrick gave a convex-dual formulation of this relation: $S(A)$ equals the maximal flux of a bounded-density flow of bulk curves from $A$ to its complement $\bar A$. These curves are called bit threads~\cite{Freedman:2016zud,Cui:2018dyq,Headrick:2017ucz,Headrick:2022nbe}.
It is tempting to view bit threads as a visualization of Bell pairs. After all, the maximal number of threads counts the entanglement entropy, as if the two endpoints of each thread could be identified with the two ends of a Bell pair. A basic difficulty is that optimal bit-thread configurations giving the same holographic entanglement entropy are not unique. This already challenges such a naive interpretation. More fundamentally, do the detailed trajectories of bit threads through the bulk have a precise quantum-information-theoretic meaning?

Our view is that this interpretational difficulty partly reflects a state-centered way of thinking. From the naturally process-centered viewpoint of categorical quantum mechanics (CQM)~\cite{Abramsky:2004doh,Abramsky:2008qkz,Coecke:2005clw} and string-diagrammatic reasoning, in particular the ZX calculus~\cite{Coecke:2008lcg,Duncan:2009ocf,vandeWetering:2020giq,Kissinger:2024pqs,Coecke:2017dti}, bit threads admit a natural interpretation in terms of quantum information flow (QIF), introduced by Coecke in Refs.~\cite{Coecke:2004sxv,Coecke:2005bin}.

More precisely, we propose the following. Consider a bipartite holographic state $\lvert\psi\rangle_{A\bar A}$ and a successful protocol that faithfully transfers up to $k$ message qubits from $A$ to $\bar A$. The QIF trajectories within the tensor-network state sector representing $\lvert\psi\rangle_{A\bar A}$ then provide an interpretation of the corresponding bit-thread configuration.
This viewpoint also gives an elegant understanding of the nonuniqueness of optimal bit-thread configurations. A fixed resource state $\lvert\psi\rangle_{A\bar A}$ can generally admit different optimal transfer protocols, despite having a fixed entanglement-limited capacity. These protocols can in turn give rise to different QIF trajectories.

To connect bit threads with such a definite microscopic process meaning, it turns out that we need a language in which an identity channel can be certified within a protocol.
Such a language is provided by CQM: quantum processes are represented by string diagrams~\cite{Joyal:1991esw}, and local semantics-preserving rewrites encode exact process equalities. We use in particular ZX string diagrams, whose rewrite calculus is especially well suited to stabilizer quantum mechanics.

Building on this string-diagrammatic language, in this Letter we show that, for holographic tensor networks with a preferred discrete geometry and finite bond dimensions, not only can QIFs attain a maximal flux matching the entanglement entropy, but, as proposed above, their trajectories themselves also obey precisely the divergence-free and density-bound conditions of Freedman-Headrick bit threads. In particular, for stabilizer-type holographic tensor networks, we give a systematic construction of such QIF configurations.
In the HaPPY model~\cite{Pastawski:2015qua}, we further construct QIF realizations of geodesic-like bit-thread configurations~\cite{Agon:2018lwq}, and find that the resulting QIF trajectories contain information beyond entanglement entropy, thereby probing finer features of the underlying entanglement structure.

\paragraph{Bit threads.}
Bit threads are usually formulated on continuous manifolds. Since our primary concern here is their quantum-information-theoretic meaning, however, the discrete tensor-network setting provides a clearer and more operational framework~\cite{note:planck-thickness}.

In holographic tensor networks (see, e.g., Refs.~\cite{Swingle:2009bg,Swingle:2012wq,Pastawski:2015qua,Hayden:2016cfa,Bao:2018pvs}), the bulk geometry is discretized by a graph $\mathcal{G}=(V,E,c)$. Each edge $e$ is associated with a bond Hilbert space of dimension $\chi_e$, and we define its capacity as $c_e=\log_2 \chi_e .$ 
A cut $C\sim A$ separating the boundary region $A$ from its complement $\bar A$ is a set of edges whose removal separates the two regions. The total capacity $\sum_{e\in C}c_e$ plays the role of the length of a curve in the continuum geometry. Thus, in units of bits, the RT formula for a holographic tensor network takes the form
\begin{equation}
S(A)=\min_{C\sim A}\sum_{e\in C}c_e,
\label{eq:discreteRT}
\end{equation}
where convention-dependent normalization factors have been suppressed.

A discrete formulation of bit threads was given in Ref.~\cite{Cui:2018dyq}. A discrete bit-thread configuration $\mathcal B$ is a family of network paths connecting $A$ to $\bar A$. The paths cannot begin or end in the bulk, and the number $n_e(\mathcal B)$ of threads passing through any bond $e$ cannot exceed its capacity:
\begin{equation}
n_e(\mathcal B)\leq c_e .
\label{den}
\end{equation}
If every path is oriented from $A$ to $\bar A$, the requirement that no thread begins or ends in the bulk is equivalently expressed as flow conservation at every bulk vertex $v$, or
\begin{equation}
\sum_{e\in \mathrm{out}(v)} n_e(\mathcal B)
=
\sum_{e\in \mathrm{in}(v)} n_e(\mathcal B) .
\label{div}
\end{equation}
These are the network counterparts of the density bound and the divergenceless condition for continuum bit threads, respectively. The max flow-min cut theorem further guarantees that the maximal flux of an admissible bit-thread configuration equals the capacity of the minimal cut. 
The discrete formulation does not change the essential content of the bit-thread construction, but places it in a setting better suited to comparison with QIFs in string diagrams.

\begin{figure}[t]
    \centering
    \includegraphics[width=\columnwidth]{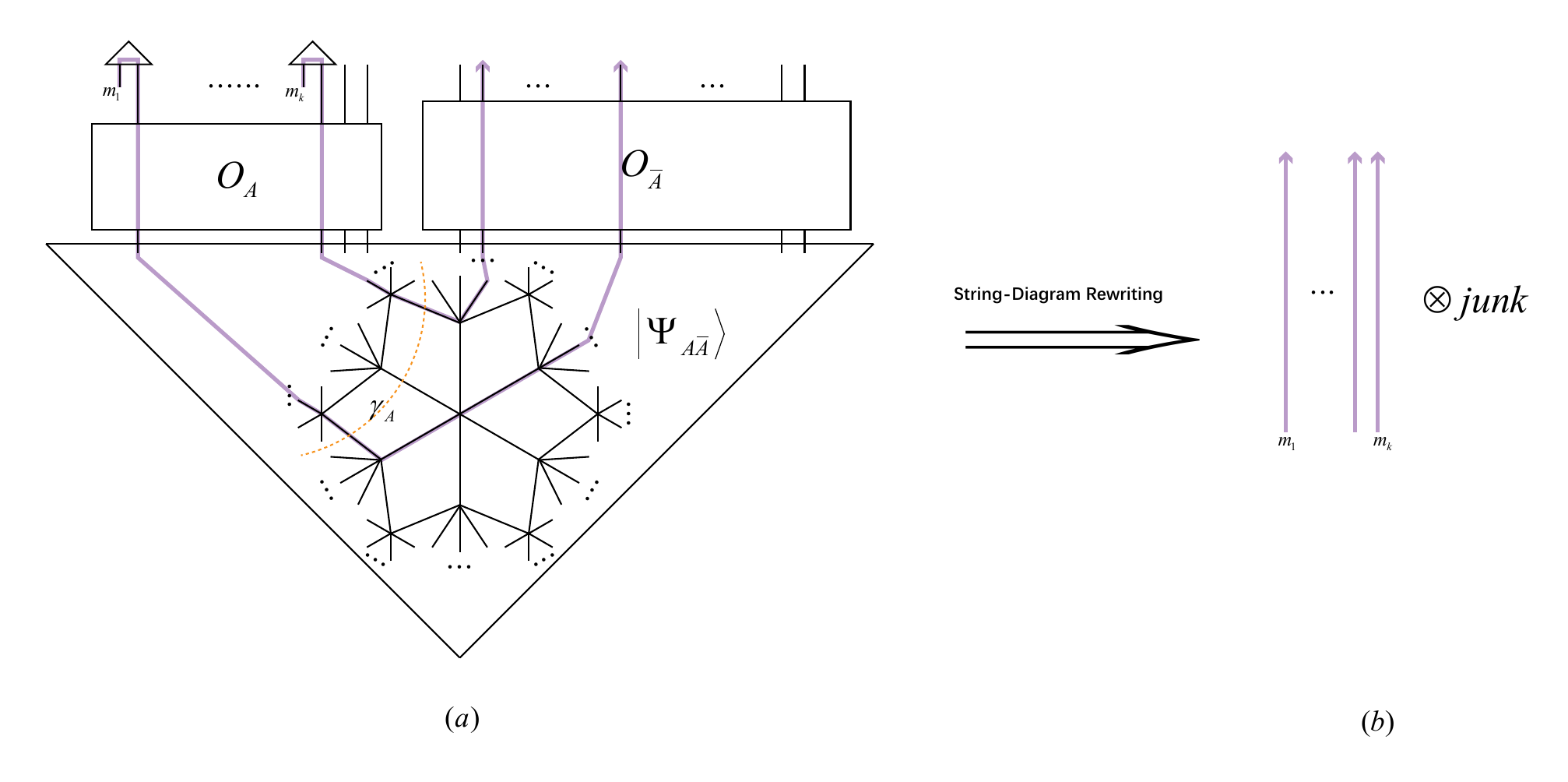}
   \caption{
QIF picture in a holographic protocol network.
(a) The entanglement-network string diagram represents an LOCC protocol in which operations $O_A$ and $O_{\bar A}$ are first applied to the two boundary sides of the bipartite holographic tensor-network state $\ket{\psi_{A\bar A}}$, distilling as many Bell pairs as possible, after which standard teleportation is used to transfer the qubit messages $m_1,\ldots,m_k$. Due to holographic duality, the network structure of $\ket{\psi_{A\bar A}}$ corresponds to a preferred semiclassical geometry, wherein RT surface $\gamma_A$ is depicted as the orange dashed line. The QIFs characterizing the transfer of the $k$ qubits are marked in purple. We argue that the portions of their trajectories passing through the tensor-network section precisely satisfy the bit-thread constraints.
(b) For each measurement branch, the protocol string diagram can be rewritten, while preserving its morphism semantics, into a representation that is as simple as possible. Ultimately, the $k$ QIFs are exhibited in this simplified representation by $k$ clean bare wires.}
    \label{fig-main}
\end{figure}

\paragraph{Quantum information flow.}
\begin{figure}[t]
    \centering
    \includegraphics[width=\columnwidth]{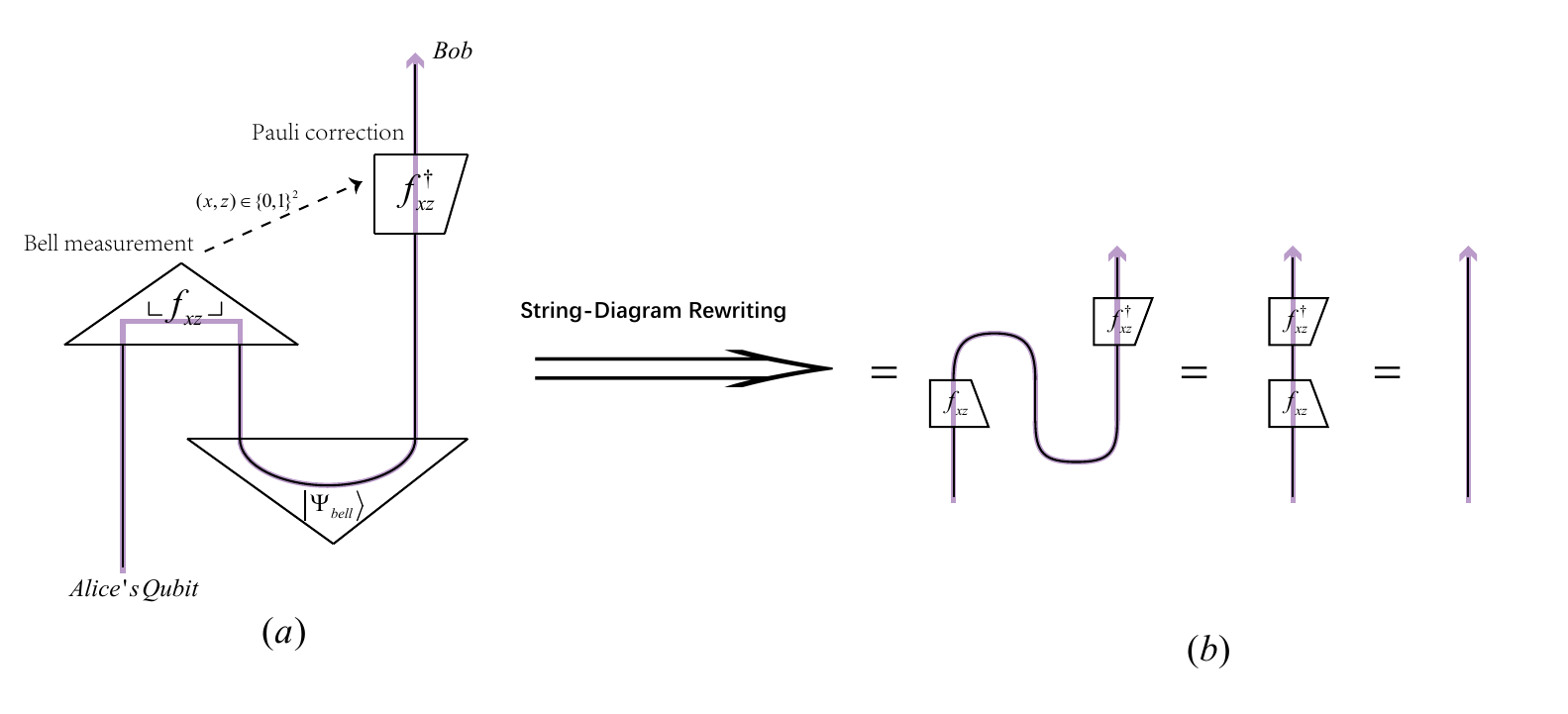}
    \caption{  QIF in protocol-network geometry.  (a) String diagram of the teleportation protocol network, with a set of measurement outcomes $\{x,z\}$ fixed and the parties' operations coordinated accordingly, termed an \emph{entanglement network} in~\cite{Coecke:2004sxv,Coecke:2005bin}. The black dashed lines denote the usual classical communication. The purple line represents the QIF supported on the protocol-network string diagram. (b) String-diagram rewriting for teleportation. }
    \label{fig-tele}
\end{figure}
The notion of QIF originates in Bob Coecke's reinterpretation of well-known protocols such as quantum teleportation from the perspective of categorical process theory~\cite{Coecke:2004sxv,Coecke:2005bin}.
The basic idea is as follows. For a quantum protocol involving measurements, one can speak of classical information flow, in which measurement outcomes are communicated classically to coordinate the parties' operations. Once a particular set of measurement outcomes is fixed, one can also define, within the resulting protocol network, a path constrained by specific local rules. In a certain sense, such a path reveals the mechanism of information transfer within the protocol, and is therefore called a quantum information flow (QIF).
Figure~\ref{fig-tele}(a) shows the protocol network and the QIF supported on it for the simple example of quantum teleportation~\cite{Bennett:1992tv,Nielsen:2012yss}.
More precisely, it is the standard string-diagram representation of the corresponding protocol network in CQM. Categorically, wires denote objects and boxes denote morphisms; physically, they represent quantum systems and quantum processes, respectively. Vertical composition and horizontal juxtaposition represent composition and tensor product. Reading the diagram from bottom to top, a triangular box with no input represents a quantum state, i.e., a ket, whereas a triangular box with no output represents a bra specifying a measurement branch.
The spirit of CQM~\cite{Abramsky:2004doh,Abramsky:2008qkz,Coecke:2005clw} is that, once quantum processes are cast into a standard string-diagrammatic form grounded in categorical semantics, local diagrammatic rewrites can be performed while preserving the morphism represented by the diagram as a whole. Such rewrites can expose a simpler structure hidden in an apparently complicated protocol. In particular, a bare straight wire represents an identity morphism, meaning an identity channel physically. 
A QIF is such a through-path on an entanglement network that, under each step of string-diagram rewriting, can be compatibly inherited to the next diagram, until it is finally displayed transparently on a clean bare wire. The same requirement must hold for every measurement branch. See Fig.~\ref{fig-tele}(b).
For our purposes, we are interested in a particular class of protocol networks, as illustrated in Fig.~\ref{fig-main}, in which the entangled resource state is a nontrivial holographic tensor-network state with a preferred physical geometry, and the two parties perform suitable operations on it so as to transfer as many quantum messages as possible. A successful transfer of $k$ message qubits then gives rise to $k$ QIFs in the protocol-network geometry. The portions of these QIF trajectories that pass through the holographic state sector are precisely what we will compare with bit threads.


\paragraph{Capacity correspondence.}

Consider a bipartite state $\psi_{A\bar A}$ given by a holographic tensor-network model satisfying the RT formula, and LOCC operations between $A$ and $\bar A$. 
At the leading order relevant to the present work, and assuming that the relevant errors are appropriately controlled, the holographic entanglement entropy counts the number of Bell pairs obtainable by optimal one-shot entanglement distillation~\cite{note2}.
Since each Bell pair can be used, by standard quantum teleportation, to transfer one message qubit along a corresponding QIF, therefore, denoting the number of qubits transferred from $A$ to $\bar A$ by a protocol $\mathcal P$ as $F_q\!\left(A\rightarrow\bar A\mid\psi_{A\bar A},\mathcal P\right)$, then
\begin{equation}
S(A)
=
\max_{\mathcal P}
F_q\!\left(A\rightarrow\bar A\mid\psi_{A\bar A},\mathcal P\right),
\label{coe}
\end{equation}
where $\mathcal P$ ranges over all allowed protocols. In this sense, the holographic entanglement entropy gives the maximal QIF capacity. Fig.~\ref{fig-main}(a) shows the entanglement-network string diagram corresponding to such an optimal protocol $\mathcal P$.
By comparison, in the bit-thread formulation of the RT formula, the same entropy is given by the maximal bit-thread flux $F_b(A\rightarrow\bar A\mid\mathcal B)$:
\begin{equation}
S(A)
=
\max_{\mathcal B}
F_b\!\left(A\rightarrow\bar A\mid\mathcal B\right),
\label{bit}
\end{equation}
where $\mathcal B$ ranges over all admissible bit-thread configurations satisfying the divergenceless condition and the density bound. Notice the evident parallel between Eqs.~\eqref{coe} and~\eqref{bit}; convention-dependent factors have been omitted for simplicity. 
However, Eq.~\eqref{bit} is a purely mathematical consequence of the convex-programming duality associated with the minimal-area problem on the bulk geometry, whereas Eq.~\eqref{coe}  is derived from the specific entanglement structure encoded in the quantum state $\psi_{A\bar A}$.
In any case, we see that QIFs and bit threads already exhibit a natural correspondence at the level of capacity. More nontrivial, however, is their correspondence at the level of flow trajectories, to which we now turn.


\paragraph{Trajectory correspondence.}
We now show that the QIF trajectories in the protocol network for transferring quantum messages through a holographic tensor network, illustrated in Fig.~\ref{fig-main}(a), satisfy both the divergenceless condition~\eqref{div} and the density bound~\eqref{den} of Headrick-Freedman bit threads.
For this purpose, we further refine the Hilbert space associated with each macroscopic bond $e$ as $
\mathcal H_e\simeq (\mathbb C^2)^{\otimes c_e},$
corresponding to $c_e$ unit-capacity wires
\begin{equation}
B_e=\{e_{e,1},\ldots,e_{e,c_e}\},
\qquad
|B_e|=c_e .
\end{equation}
Each macroscopic tensor $v\in V$ is then expanded into a local process subdiagram $R_v$ connecting these wires. The same decomposition is applied to the operation tensors $O_A$ and $O_{\bar A}$ on the two sides.
The entire entanglement network is thereby represented by a semantics-preserving microscopic string diagram. 
Let $D_0$ denote the protocol-network string diagram, with the measurement-branch label temporarily suppressed, and write a semantics-preserving rewriting sequence as
\begin{equation}
D_0
\overset{\rho_1}{\Longrightarrow}
D_1
\overset{\rho_2}{\Longrightarrow}
\cdots
\overset{\rho_n}{\Longrightarrow}
D_n .
\label{rew0}
\end{equation}
To say that there are $k$ QIFs means that such a rewriting sequence exists for which
\begin{equation}
D_n\simeq
\left(
\bigotimes_{i=1}^{k} I_{m_i\to b_i}
\right)
\otimes R ,
\label{fin0}
\end{equation}
where, as shown in Fig.~\ref{fig-main}(b), $I_{m_i\to b_i}$ is a bare identity wire from an incoming wire $m_i$ to an outgoing wire $b_i$, while $R$ denotes the remaining junk diagram decoupled from these wires.
For the protocols involving only simple bipartite entanglement considered in Coecke's original discussions~\cite{Coecke:2004sxv,Coecke:2005bin}, the rewriting sequence~\eqref{rew0} is usually rather short. As a result, the QIF trajectories in $D_0$ can be identified almost directly from the terminal form~\eqref{fin0}, and can be described by assigning local traversal rules to the elementary component boxes.
However, general multipartite entanglement and sufficiently complicated rewriting sequences both call for a more careful formalization. 
Such a rigorous formalization is provided in Ref.~\cite{Lin:2026hpd}. There, for the entanglement-network string diagram $D_0$ of a deterministic protocol, given Eqs.~\eqref{rew0} and~\eqref{fin0}, one may in principle start from the trivial through-path $\Gamma_r^{(n)}$ supported on the bare identity wire $I_{m_r\to b_r}$ in $D_n$, and trace it backward step by step through the rewriting sequence using the compatible-inheritance relation $\mathcal T_{\rho_k}$ associated with each rewrite $\rho_k$. This procedure yields, in $D_0$, the desired branch-independent constrained QIF trajectory $\Gamma_r^{(0)}$ ~\cite{note3}.
In particular, $\Gamma_r^{(0)}$ is an apparent through-path in $D_0$, i.e., it never cross any visible tensor-product gap in $D_0$. More strongly, because it admits compatible representatives throughout the entire rewriting sequence and is ultimately inherited onto the bare wire in Eq.~\eqref{fin0}, it has genuine through-goingness in the sense of Ref~\cite{Lin:2026hpd}.

In what follows, we suppress the superscript $(0)$ and denote by
\begin{equation}
\mathcal F=\{\Gamma_r\}_{r=1}^{k}
\end{equation}
the family of QIF trajectories, oriented from $A$ to $\bar A$, on the microscopic refinement of Fig.~\ref{fig-main}(a), and we will restrict attention to the portions of these trajectories lying within the holographic tensor network sector.
For each macroscopic tensor vertex $v$, whenever a trajectory $\Gamma_r$ passes through a microscopic process node $a$ in $v$’s string-diagram presentation $R_v$, its through-goingness implies that it must enter $a$ along one incident wire and leave along another. 
Thus each trajectory already satisfies flow conservation at the level of microscopic nodes. Summing these microscopic conservation relations over $R_v$ then yields flow conservation at every bulk macroscopic vertex,
\begin{equation}
\sum_{e\in\mathrm{out}(v)} n_e(\mathcal F)
=
\sum_{e\in\mathrm{in}(v)} n_e(\mathcal F),
\label{div2}
\end{equation}
in agreement with Eq.~\eqref{div}. Here $n_e(\mathcal F)$ denotes the number of microscopic wires in the macroscopic bond $e$ that are actually marked by the flow family.
Moreover, recall that the bundle $B_e$ associated with the bond $e$ contains exactly $c_e$ elementary qubit wires. Hence the number of microscopic wires marked by the flow family cannot exceed $c_e$:
\begin{equation}
n_e(\mathcal F)\leq c_e .
\label{den2}
\end{equation}
This is precisely the density bound~\eqref{den}.
Therefore, within the holographic tensor network sector, the QIF trajectories satisfy both the divergenceless condition and the density bound of discrete bit threads. When the number of flows saturates the capacity bound in Eq.~\eqref{coe}, the resulting configuration is a maximal bit-thread configuration.

The correspondence above also provides a process-theoretic physical interpretation of the nonuniqueness of bit-thread configurations. The same entangled resource state and the same $A\to\bar A$ transfer task can generally admit different successful implementation protocols. Different protocols have different entanglement-network string diagrams and can therefore certify different configurations of QIF trajectories~\cite{note4}. In the Supplemental Material, we illustrate this physical nonuniqueness with a simple GHZ-assisted teleportation prototype \cite{Karlsson:1998opa, Hillery:1998yq, Hillebrand:2011thesis} and with a single-layer holographic HaPPY disk state \cite{Pastawski:2015qua}.
A more formal derivation of the bit-thread constraints satisfied by QIF trajectories, together with further discussion, is given in the Supplemental Material.

\paragraph{Constructive realization in holographic stabilizer models.}
Several representative and exactly tractable holographic tensor-network models possess stabilizer structure~\cite{Pastawski:2015qua,Hayden:2016cfa,Nezami:2016zni}. For resource states of this kind, it is well known in quantum information theory that one can systematically identify, from the stabilizer generators, the Bell blocks crossing the $A|\bar A$ bipartition~\cite{Fattal:2004frh}, and explicitly construct the local Clifford operations $O_A$ and $O_{\bar A}$ in Fig.~\ref{fig-main}(a) that transform the resource state into a tensor product of Bell pairs and local junk on the two sides. The number of Bell pairs obtained in this way is exactly $S(A)$, so the QIF flux generated by the protocol in Fig.~\ref{fig-main} saturates the entanglement entropy.
Moreover, $O_A$ and $O_{\bar A}$ can be fully decomposed into elementary Clifford gates~\cite{Dehaene:2003thc,Aaronson:2004xuh}. These gates, together with the subsequent standard teleportation protocol, can all be represented within the complete stabilizer ZX calculus~\cite{vandeWetering:2020giq,Kissinger:2024pqs}. The entire protocol network can therefore be expanded into an explicit ZX string diagram, whose QIF trajectories can be certified step by step through string-diagram rewriting and compatible inheritance. The complete construction starting from the stabilizer generators is given in the Supplemental Material.

\paragraph{Geodesic-like QIFs in the HaPPY model.}

\begin{figure}[t]
    \centering
    \includegraphics[width=0.72\columnwidth]{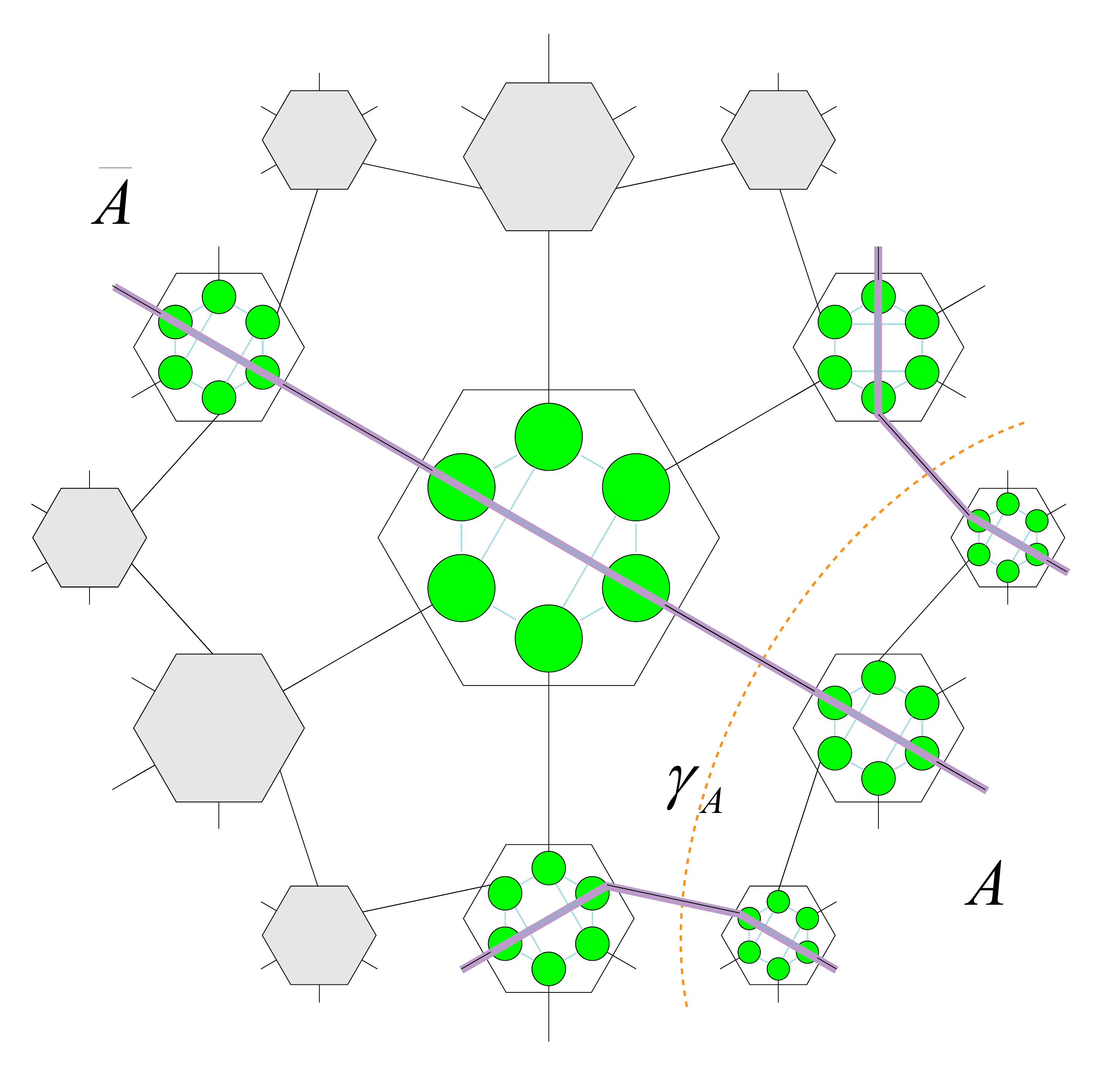}
 \caption{Geodesic-like maximal QIF in a HaPPY network. Gray hexagons denote unresolved six-leg AME perfect tensors; the remaining cells are expanded into microscopic ZX string diagrams \cite{Helwig:2013qoq, Backens:2013hto}, with green nodes denoting ZX spiders and blue dashed edges carrying Hadamard gates.
Perfectness fixes only the isometric property associated with the relevant bipartition, not the underlying state vector or its ZX representation; here we use one particular microscopic choice compatible with the target trajectories.
The purple curves mark the certified QIF trajectories, whose number saturates the capacity of the minimal cut $\gamma_A$ shown by the orange dashed curve, and therefore form a geodesic-like maximal bit-thread configuration.}  \label{fig-happ}
\end{figure}

Geometric studies of Headrick-Freedman bit threads have shown that geodesic-like configurations form a particularly representative and geometrically transparent class of explicit thread configurations~\cite{Agon:2018lwq}. If the QIF interpretation is to go beyond reproducing the maximal flux and address the geometric content of bit threads, it is therefore natural to ask whether such configurations can themselves be realized.
As shown in Fig.~\ref{fig-happ}, we consider a HaPPY disk network built from six-leg AME perfect tensors~\cite{Pastawski:2015qua}, and show that geodesic-like QIF trajectories can be rigorously certified when the microscopic ZX diagrams representing the relevant AME cells are chosen compatibly with the target trajectories, with their flux saturating the minimal cut $\gamma_A$. Technical details are given in the Supplemental Material.
It is already well known that perfectness is sufficient to determine the RT entropy and the maximal flow~\cite{Pastawski:2015qua}. What the nontrivial trajectory certification in Fig.~\ref{fig-happ} shows, however, is that specifying the flow trajectories requires information beyond perfectness, namely finer details of the underlying entanglement structure. QIF trajectories can therefore probe finer features of the entanglement structure that are invisible at the level of entropy alone.
This leaves an important open question: do all geometric bit-thread configurations admitted by the Headrick-Freedman framework admit a QIF interpretation, or can only a subset of them be lifted in this way?
A more optimistic possibility is that, for genuine holographic states, every thread configuration admitted by the former framework should acquire the physical meaning provided by the latter. If so, the allowed geometric bit thread configurations could be read in reverse as constraint data on the entanglement structure that a holographic state must be able to support.

\paragraph{Conclusion and outlook.}
Inspired by the process-centered viewpoint of categorical quantum mechanics, in this Letter we have carefully examined the quantum-information-theoretic meaning of the geometric trajectories of holographic bit threads. We have shown that, for a given holographic tensor-network state, a successful protocol for an entanglement-distillation or quantum-message-transfer task can be characterized by a protocol-geometry string diagram, and that the trajectories of the quantum information flows within this string diagram, when passing through the tensor-network sector, faithfully capture the trajectory features of bit threads. This naturally explains the long-standing nonuniqueness of optimal bit-thread configurations, while also opening a window: such line-like geometric trajectories may probe holographic quantum-entanglement structures beyond the level of entanglement entropy.It is worth emphasizing that this trajectory correspondence in fact holds at a rather general structural level: in principle, it is sufficient that the underlying holographic tensor network possess a preferred discrete geometric skeleton and finite bond dimensions.
In particular, we focus on stabilizer-type models---which are particularly amenable to reasoning with ZX string diagrams---thereby providing a systematic and diagrammatically explicit demonstration of this general correspondence.

Going even further, one may envisage using such trajectories to probe the mechanism of holographic duality itself. Here we offer some thoughts in this direction. 
When there exist effective-field-theory degrees of freedom on a fixed holographic bulk geometry, the entropy of the bulk quantum state itself also contributes to the entanglement entropy of the boundary state, appearing as a quantum correction to the RT formula~\cite{Faulkner:2013ana,Engelhardt:2014gca}. The holographic quantum-error-correction picture suggests an encoding description in which these bulk EFT degrees are regarded as logical qubits encoded into the boundary CFT, schematically,
$
|\Psi\rangle_{\partial}={\mathcal{V}}|\psi\rangle_{\rm bulk},
$
with the state vector $|\psi\rangle_{\rm bulk}$ represented by additional internal legs of the tensor network---the ``logical legs''~\cite{Almheiri:2014lwa,Pastawski:2015qua}. CQM then again offers a further clue: a state vector is itself a process, or mathematically a morphism. One may therefore resolve the bulk state into a nontrivial preparation circuit,
$
|\psi\rangle_{\rm bulk}={\mathcal{U}}|0\cdots0\rangle .
$
It is then natural to consider, in the same way, a boundary distillation or bipartite-transfer protocol on the two-layer process diagram ${\mathcal{VU}}$, and to examine the QIFs in the complete protocol string diagram. It is not difficult to see that a QIF may now either travel continuously along the tensor-network slice ${\mathcal{V}}$, or flow ``upward’’ through ${\mathcal{U}}$ midway and subsequently return to the slice ${\mathcal{V}}$. 
If we look only at the tensor-network slice ${\mathcal{V}}$, however, this leads to the following picture: some lines remain uninterrupted throughout, whereas others terminate at some point on the slice ${\mathcal{V}}$, forming a sink, and then reappear elsewhere, forming a source. Actually, this faithfully captures the behavior of the so-called quantum bit threads~\cite{Agon:2021tia,Rolph:2021hgz,Headrick:2025awv}---a variant of bit threads likewise developed through convex-optimization methods to describe the quantum-corrected RT formula. From the viewpoint of QIF, however, the two are in this sense unified. We leave the study of QIF in quantum bit threads and in deeper quantum-error-correction mechanisms to future work.

Another point worth emphasizing is that our use of the ZX calculus to study stabilizer-type holographic tensor-network models serves only to obtain controlled, explicit, and rigorous constructions. 
The notion of QIF itself is not tied to any particular specialized string-diagrammatic calculus or the categorical structure embedded in it~\cite{notex}. 
A meaningful next step is therefore to examine the proposal of this Letter and related developments in more realistic holographic tensor-network models---where by ``more realistic'' we mean models that more closely conform to or reflect the genuine microscopic structure of the CFT itself, as in  e.g.~\cite{Chen:2022wvy,Hung:2024gma,Hung:2025vgs,Geng:2025efs}.

\paragraph*{Acknowledgement}
We would like to thank Chen-Ye Li for valuable discussions. This work is supported by the NSFC Grant No.1250050230.

\appendix
\setcounter{secnumdepth}{1}

\section{CQM and ZX string-diagram preliminaries}\label{appa}

\begin{figure}[t]
    \centering
    \includegraphics[width=\columnwidth]{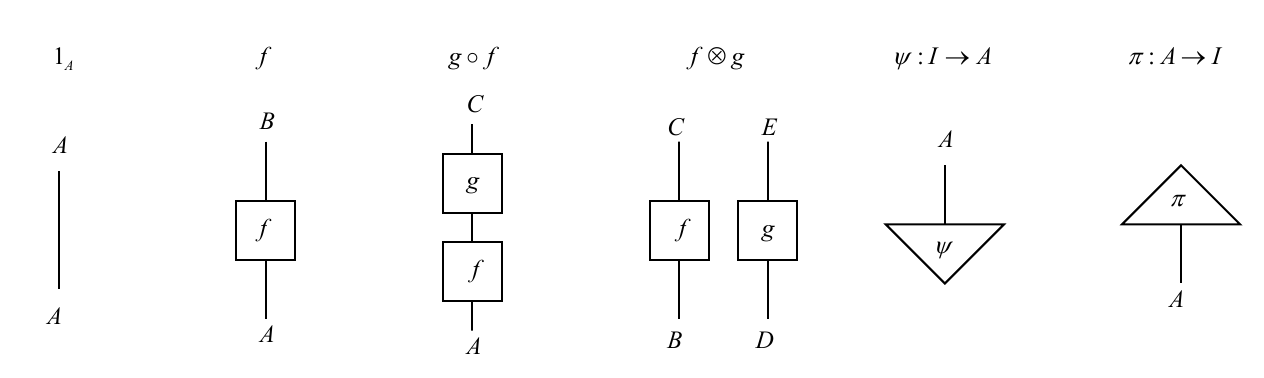}
    \caption{ Graphical representations of various morphisms in string diagrams.}
    \label{fig-stri}
\end{figure}

In this paper, we use the string-diagram language of categorical quantum mechanics to represent quantum states, quantum operations, and their composition, and employ the ZX calculus as a concrete rewriting tool in the stabilizer setting. Here we review only the minimal structure needed below. For a systematic development of this language, a complete string-diagram derivation of quantum teleportation, and a general formalization of quantum information flow, see Ref.~ \cite{Lin:2026hpd}.
In a string diagram, a wire represents the type of a quantum system, while a box represents a quantum process. A morphism 
$
f:A\longrightarrow B
$
is drawn as a box with an input wire $A$ and an output wire $B$. Connecting two processes sequentially along a wire represents their composition, whereas placing two diagrams side by side represents the tensor product of the corresponding processes. A string diagram is therefore not merely a schematic drawing of a quantum circuit: under the corresponding monoidal-categorical semantics, the way in which the diagram is connected directly expresses the compositional structure of the processes.
States and effects are special cases of this syntax. A state (i.e., Dirac ket)
$
\psi:I\longrightarrow A
$
represents a process that prepares the system $A$ from the tensor unit $I$, while an effect (i.e., Dirac bra)
$
\pi:A\longrightarrow I
$
takes $A$ as input and leaves no quantum output. Throughout this work, string diagrams are read from bottom to top. Thus, the wires of a state extend upward from its graphical representation, whereas the wires of an effect enter its graphical representation from below. See Fig.~\ref{fig-stri}.

\begin{figure}[t]
    \centering
    \includegraphics[width=\columnwidth]{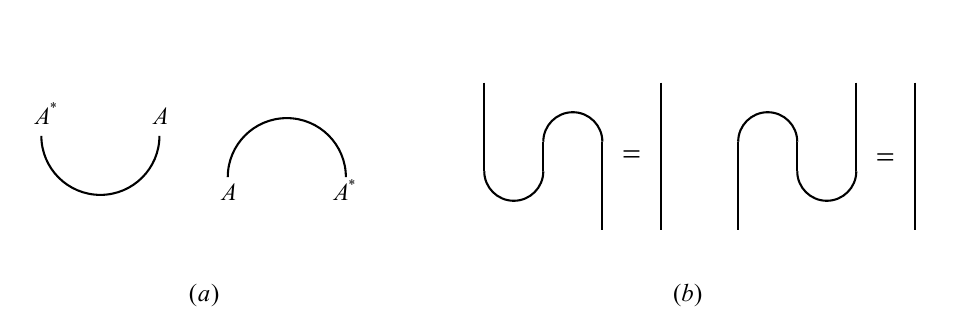}
    \caption{ (a) The cup and the cap. 
(b) The snake equations. }
    \label{fig-comp}
\end{figure}

For finite-dimensional pure-process quantum theory, one may in addition use a dagger-compact structure. For each system $A$, this provides a pair of cup and cap morphisms,
\begin{equation}
\eta_A:I\longrightarrow A^*\otimes A,
\qquad
\varepsilon_A:A\otimes A^*\longrightarrow I,
\end{equation}
which in particular satisfy the snake identities. Graphically, these identities allow a bent wire to be continuously straightened, as illustrated in Fig.~\ref{fig-comp}. In the quantum-information setting, the qubit cup is usually taken to be an unnormalized Bell state, while the corresponding cap represents a Bell-type effect.
The CQM proof of quantum teleportation is a standard example reflecting precisely this snake-equation structure. For each fixed branch of measurement outcomes, the apparently complicated process diagram consisting of a Bell resource, a Bell-type effect, and branch-conditioned corrections can be reduced, by local string-diagram equalities, to
$
\operatorname{id}_{m\to b},
$
namely, a bare identity wire from the message input $m$ to the output $b$. Its physical meaning is the identity quantum process from $m$ to $b$.
More generally, a quantum protocol may contain measurements, in which case we consider each fixed branch of classical outcomes separately. Once a branch is fixed, the operations conditioned on the measurement outcomes become definite morphisms, and the corresponding branch-fixed protocol diagram becomes an ordinary pure-process string diagram. Deterministic success of the protocol requires that, for every classical-outcome branch, the resulting diagram can be reduced to the same
$
\operatorname{id}_{m\to b}.
$
This is also a prerequisite for speaking of quantum information flow.

\begin{figure}[t]
    \centering
    \includegraphics[width=\columnwidth]{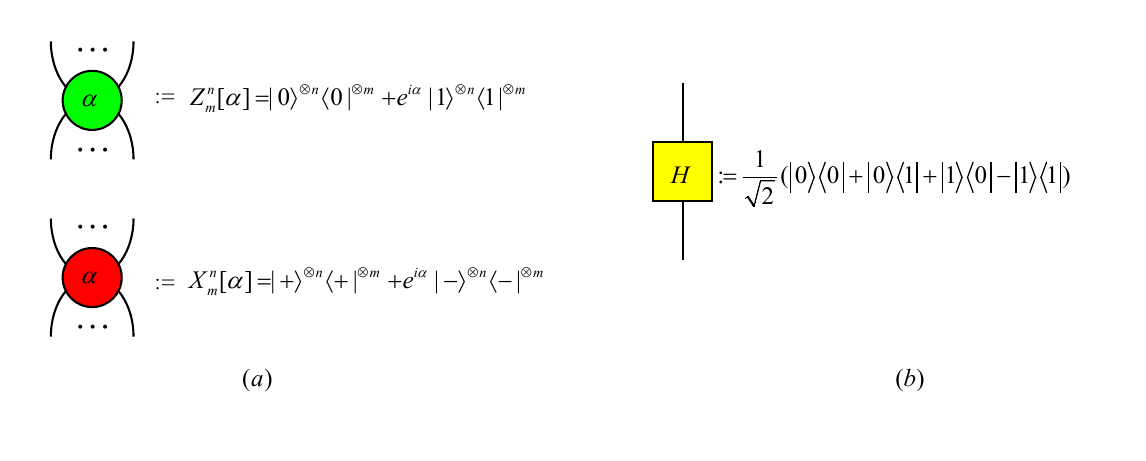}
    \caption{    Basic generators of the ZX calculus.     }
    \label{fig-spid}
\end{figure}

To carry out explicit local simplifications of the stabilizer states and Clifford protocols considered in this paper, we now use the ZX calculus as a concrete realization of the general string-diagram language above.
A ZX diagram is not a separate graphical language detached from the preceding CQM setting. 
It is built on the same dagger-compact string-diagrammatic framework, but makes this framework concrete for qubit processes by choosing a specific set of generators and rewrite rules. The generators are the $Z$-spider, the $X$-spider, and the Hadamard gate, shown in Fig.~\ref{fig-spid}. 
We retain the convention that diagrams are read from bottom to top.
The Hadamard gate is represented simply by a yellow box. A green $Z$-spider with $m$ input wires, $n$ output wires, and phase $\alpha$ represents the linear map
\begin{equation}
Z_{\alpha}^{m,n}
=
|0\rangle^{\otimes n}\langle 0|^{\otimes m}
+
e^{i\alpha}
|1\rangle^{\otimes n}\langle 1|^{\otimes m}.
\label{zspid}
\end{equation}
The corresponding red $X$-spider is defined by
\begin{equation}
X_{\alpha}^{m,n}
=
|+\rangle^{\otimes n}\langle +|^{\otimes m}
+
e^{i\alpha}
|-\rangle^{\otimes n}\langle -|^{\otimes m},
\label{xspid}
\end{equation}
where
$
|\pm\rangle=\frac{|0\rangle\pm|1\rangle}{\sqrt{2}}.
$
When $\alpha=0$, the phase label is usually omitted. We also allow $m=0$ or $n=0$: the former gives a state, while the latter gives an effect.

\begin{figure}[t]
    \centering
    \includegraphics[width=\columnwidth]{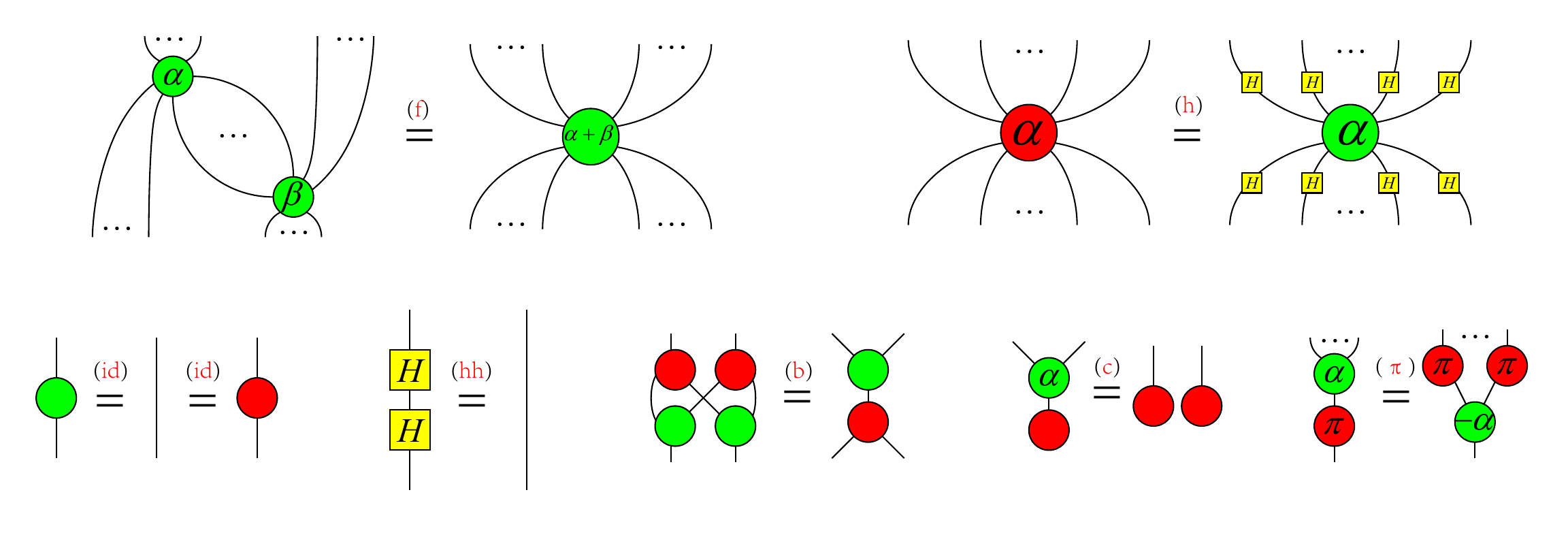}
    \caption{ Complete rule set for Clifford ZX-diagrams. Equality is understood up to a nonzero scalar. }
    \label{fig-rule}
\end{figure}

For the Clifford setting considered in this paper, Fig.~\ref{fig-rule} displays the basic rewrite rules of the ZX calculus~\cite{Backens:2013hto,Backens:2015nhm}. Each rule expresses an exact equality between linear maps of qubit Hilbert spaces. 
Consequently, each step in a sequence of ZX rewrites $D_0\Rightarrow D_1\Rightarrow\cdots\Rightarrow D_N$
 leaves unchanged the morphism denoted by the entire diagram.
Although the phase-dependent equalities in Fig.~\ref{fig-rule} hold for arbitrary
$
\alpha,\beta\in[0,2\pi),
$
the resource states and protocol diagrams used explicitly in this work belong to the Clifford fragment, for which spider phases are integer multiples of $\pi/2$.

\begin{figure}[t]
    \centering
    \includegraphics[width=0.52\columnwidth]{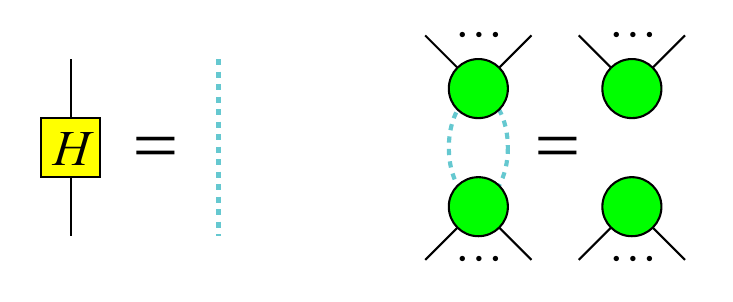}
    \caption{Hopf law.}
    \label{fig-hopf}
\end{figure}

Another relation that will be used repeatedly is the well-known Hopf law, which can be derivable from rules (b) and (c). See Fig.~\ref{fig-hopf}, where we follow a common convention in the literature in which an edge carrying a Hadamard gate is represented as a blue dashed edge. 
The Hopf law states that a pair of parallel edges connecting spiders of opposite colors can be eliminated. Together with rule (h), this equivalently implies that a pair of parallel blue edges between two green spiders can be eliminated. The significance of the Hopf law is that it reflects the strong complementarity of the two Frobenius structures underlying the ZX calculus.

For a fuller account of the axiomatic foundations of the ZX calculus, its completeness theorems, and its broader conceptual significance within categorical quantum mechanics, see Ref. \cite{Lin:2026hpd}  and the references therein.

\section{A minimal example of trajectory nonuniqueness}\label{appb}

For the holography community, it is almost a natural temptation to think of bit threads intuitively as some kind of visualization of Bell pairs, whether explicitly or implicitly (see e.g.,~\cite{Freedman:2016zud, Harper:2022sky, Bao:2023til, Lin:2022flo}): after all, the number of threads really does count entanglement entropy. Terms such as Bell pairs, distillation, and LOCC frequently appear in the relevant literature. 
Nevertheless, this very attempt to explain finds itself in an embarrassing predicament. Once we try to understand each bit thread simply as a Bell pair, we are naturally led to interpret its two endpoints as the two ends of that Bell pair. In this way, however, we immediately face a difficulty: since optimal bit-thread configurations that compute the same holographic entanglement entropy exist non-uniquely, how should one interpret the resulting indeterminacy in the positions of the endpoints of these Bell pairs? Or is it inappropriate to speak of their positions at all? More nontrivially, what physical interpretation should then be assigned to the trajectory of a bit thread inside the bulk slice?

Our judgment is that the root of this difficulty lies neither in the fact that ``bit-thread configurations are non-unique'', nor in the interpretive direction that ``bit threads are related to Bell pairs''. Rather, in a certain sense, this is because previous interpretations have remained accustomed to a state-centred perspective. More precisely, we are used to asking: why does the same resource state correspond to many different bit-thread configurations? 
By contrast, the categorical quantum mechanics and string-diagrammatic calculus are naturally to ask process-centred question: 
Given the same resource state and the same transfer task, can there be different concrete protocols that all implement the task successfully? In general, there can. Each such protocol has its own entanglement-network string diagram and can certify its own QIFs. The resulting nonuniqueness is therefore no longer mysterious: different QIF trajectories can arise from different successful implementations of the same task. When attention is restricted to the resource-state sector, the nonuniqueness of the trajectories supported there appears precisely in the form expected of nonunique bit-thread configurations.

\begin{figure}[t]
    \centering
    \includegraphics[width=\columnwidth]{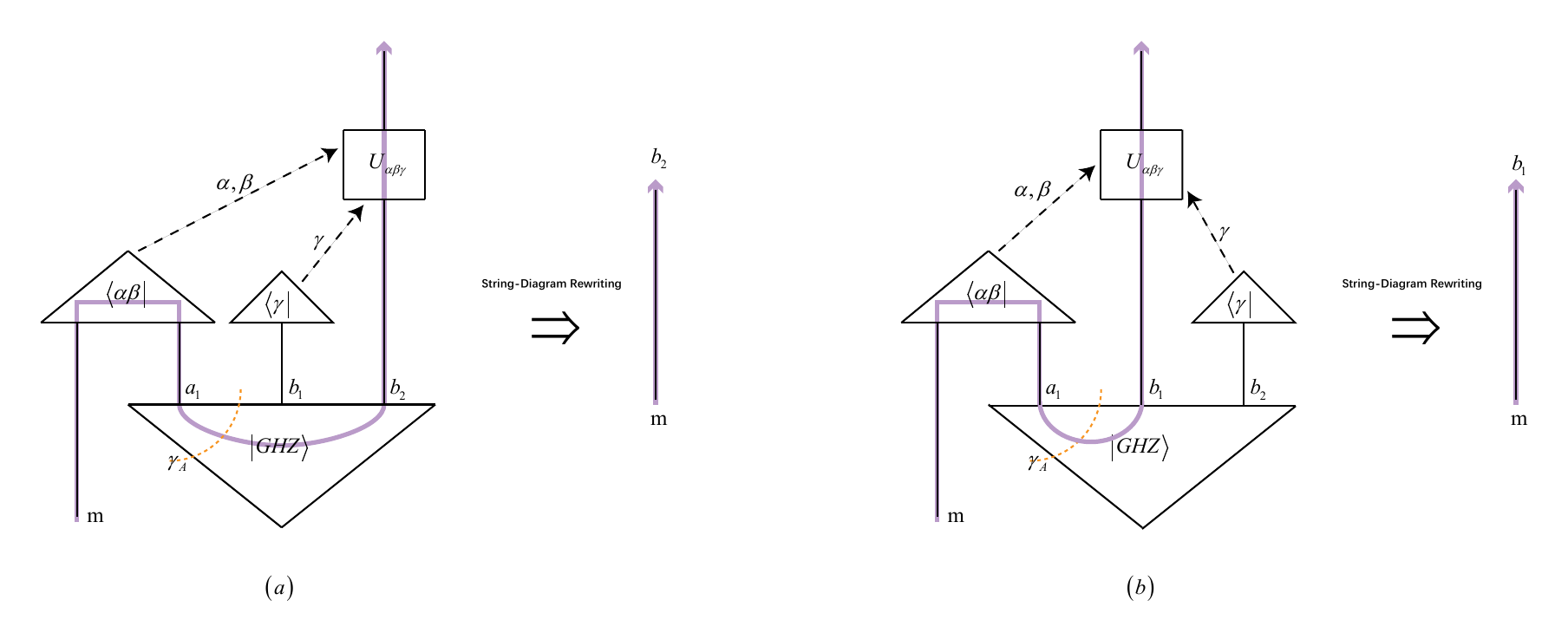}
  \caption{Two implementations of GHZ-assisted teleportation with the same resource state and transfer task. The purple lines indicate the corresponding QIF trajectories.}
    \label{fig-two}
\end{figure}

Here we present a minimal example, GHZ-assisted teleportation, to illustrate our argument.
As shown in Fig.~\ref{fig-two}, consider a three-leg GHZ resource state whose three boundary legs represent three experimenters: $a_1$ belongs to laboratory $A$, while $b_1$ and $b_2$ belong to laboratory $B$. Laboratory $A$ additionally holds a message qubit $m_1$ to be transferred. The task of interest is not to deliver $m_1$ to a particular experimenter specified in advance, but simply to transfer it successfully from laboratory $A$ to laboratory $B$. Thus, whether the message is ultimately received by $b_1$ or by $b_2$, the same $A\to B$ task is successfully accomplished. This provides a minimal prototype showing that, for the same resource state and the same task, different implementations can yield different QIF witnesses whose trajectories on the resource-state slice exhibit precisely the kind of nonuniqueness expected of bit threads.

We retain below only the minimal technical ingredients needed to compare the two trajectories. For full details of the GHZ-assisted teleportation protocol, including the complete correspondence between measurement outcomes and local corrections and the branch-by-branch reduction of the corresponding ZX string diagrams, see Refs.~\cite{Lin:2026hpd,Hillebrand:2011thesis}.

In the first protocol, shown in Fig.~\ref{fig-two}(a), $b_2$ serves as the receiving system. Laboratory $A$ performs a Bell-basis measurement on $m_1$ and $a_1$; the experimenter holding $b_1$ performs an $X$-basis measurement on the corresponding GHZ qubit; and finally $b_2$ undergoes the appropriate local Pauli correction determined by the classical outcomes of these two measurements. For every fixed branch of classical outcomes, the corresponding ZX string diagram reduces to
\begin{equation}
\operatorname{id}_{m_1\to b_2}.
\label{m1b2}
\end{equation}

The second protocol is obtained by exchanging the operational roles of $b_1$ and $b_2$: $b_2$ is now measured in the $X$ basis, while the appropriate Pauli correction, conditioned on the classical outcomes, is applied to $b_1$.
 Accordingly, every fixed branch reduces to
\begin{equation}
\operatorname{id}_{m_1\to b_1}.
\label{m1b1}
\end{equation}

The QIF trajectories associated with the two protocols can then be determined, and are marked by the purple lines in Fig.~\ref{fig-two}. We are particularly interested in the portions of these trajectories lying within the resource-state sector. The GHZ state itself provides a minimal nontrivial tensor network, consisting of a single bulk tensor with three boundary legs. The bulk trajectory obtained from the first protocol is
\begin{equation}
a_1\longrightarrow b_2 ,
\label{a1b2}
\end{equation}
whereas that obtained from the second protocol is
\begin{equation}
a_1\longrightarrow b_1 .
\label{a1b1}
\end{equation}

The two protocols therefore use the same GHZ resource state, accomplish the same $A\to B$ single-qubit transfer task, and have the same transfer capacity, while being characterized by different quantum information flows. This suggests that the nonuniqueness of bit-thread trajectories can reflect the nonuniqueness of the implementation process itself under a fixed resource state and a fixed task. Different bit-thread configurations may then be understood as the resource-state-sector manifestations of QIFs belonging to entanglement networks that characterize different concrete implementations of the same task.

\section{ZX analysis of the holographic HaPPY model}
\label{appc}

\subsection{The AME$(6,2)$ state and a ZX derivation of the perfect-tensor relation}\label{appc1}

\begin{figure}[t]
    \centering
    \includegraphics[width=0.52\columnwidth]{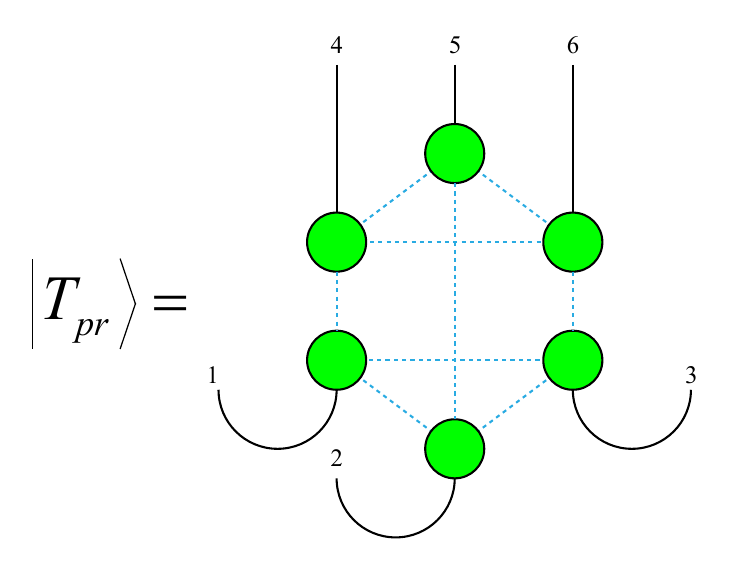}
  \caption{Triangular-prism ZX representation of an AME$(6,2)$ state. Blue dashed edges carry Hadamard gates.}    \label{fig-tpr}
\end{figure}

We consider a HaPPY disk tensor-network model~\cite{Pastawski:2015qua}, in particular the $\{6,4\}$ tiling built from AME$(6,2)$ states, and for the moment omit bulk logical legs. This is one of the canonical toy models of holographic tensor networks. Despite its simplicity--and hence its considerable advantage in tractability--it has provided valuable insight into several central ideas of holography, including the RT formula, quantum error correction, and entanglement-wedge reconstruction~\cite{Almheiri:2014lwa,Dong:2016eik,Harlow:2018fse,Pastawski:2015qua}.
The best-known property of the HaPPY model is the perfectness of its cell tensors. In the particular model considered here, each six-leg cell corresponds to an AME$(6,2)$ state
$
|\mathrm{AME}_{6,2}\rangle\in(\mathbb C^2)^{\otimes 6}.
$
Through the Choi—Jamiolkowski correspondence~\cite{Choi:1975nug, Jamiolkowski:1972pzh}, it may be regarded as a map
$
T:\mathcal H_{\rm in}\longrightarrow\mathcal H_{\rm out}
$
from any chosen set of $k\leq 3$ legs to the remaining $6-k$ legs. Perfectness then means
\begin{equation}
T^\dagger T=I_k ,
\label{per}
\end{equation}
where $I_k$ denotes the identity map on the $k$ input legs. In particular, for $k=3$, $T$ is unitary.
Using this perfect-tensor property, the HaPPY model realizes a network version of the RT formula: the entanglement entropy between $A$ and ${\bar A}$ is given precisely by the minimal cut through the network. The same perfect-tensor property also allows the quantum-error-correcting structure and entanglement-wedge reconstruction of the model to be realized explicitly~\cite{Almheiri:2014lwa,Dong:2016eik}.
It should be acknowledged, however, that at the microscopic level a tensor-network cell must ultimately correspond to a specific state vector, which we may write in Dirac notation as
\begin{equation}
|T\rangle
=
\sum_{i_1,\ldots,i_6}
T_{i_1\cdots i_6}
|i_1\cdots i_6\rangle .
\end{equation}
This microscopic specification was not made explicit in Ref.~\cite{Pastawski:2015qua}; more precisely, it was quite reasonably left unspecified. The reason is that the perfect-tensor condition--or, in the present six-qubit case, the requirement that the cell be an AME$(6,2)$ state--does not uniquely determine the state vector. For the arguments developed in Ref.~\cite{Pastawski:2015qua}, this ambiguity is not essential, since those arguments rely almost entirely on the perfect-tensor property itself.
Accordingly, for the purposes of the present analysis, we are free to choose a concrete microscopic realization. We shall therefore take the state vector associated with each six-leg cell of the HaPPY model to be represented by the ZX diagram shown in Fig.~\ref{fig-tpr}, commonly referred to as the triangular-prism representation of AME$(6,2)$~\cite{Helwig:2013qoq}. This is one perfectly admissible choice satisfying the perfect-tensor condition. In the computational basis, it may be written explicitly as
\begin{equation}
|T_{\rm pr}\rangle
=
\frac{1}{8}
\sum_{x_1,\ldots,x_6\in\{0,1\}}
(-1)^{q(x_1,\ldots,x_6)}
|x_1x_2x_3x_4x_5x_6\rangle ,
\end{equation}
with phase polynomial
\begin{equation}
\begin{split}
q(x_1,\ldots,x_6)
={}&
x_1x_2+x_2x_3+x_3x_1
+x_4x_5+x_5x_6+x_6x_4
\\
&+x_1x_4+x_2x_5+x_3x_6
\pmod 2 .
\end{split}
\end{equation}
Note that in Fig.~\ref{fig-tpr}, a blue dashed edge denotes an edge carrying a Hadamard gate.
Moreover, once the tensor network is treated as a genuine typed string diagram, wire orientation becomes part of the mathematical data rather than a merely pictorial convention. When all six legs are directed upward, the six-spider diagram represents a quantum state. If instead three of the legs are designated as inputs and directed downward, while the remaining three are directed upward as outputs, the same underlying diagram represents a $3\to3$ map, i.e., a physical quantum process. These two readings are related precisely by the Choi-Jamiolkowski correspondence in CQM~\cite{Abramsky:2004doh,Abramsky:2008qkz,Coecke:2005clw}. In tensor-network practice, this relation is often used implicitly, without the distinction being made explicit~\footnote{However, Refs. like~\cite{Wood:2011zvw, Biamonte:2011aoz, Meznaric:2012yae} are the exceptions.}.

\begin{figure}[t]
    \centering
    \includegraphics[width=\columnwidth]{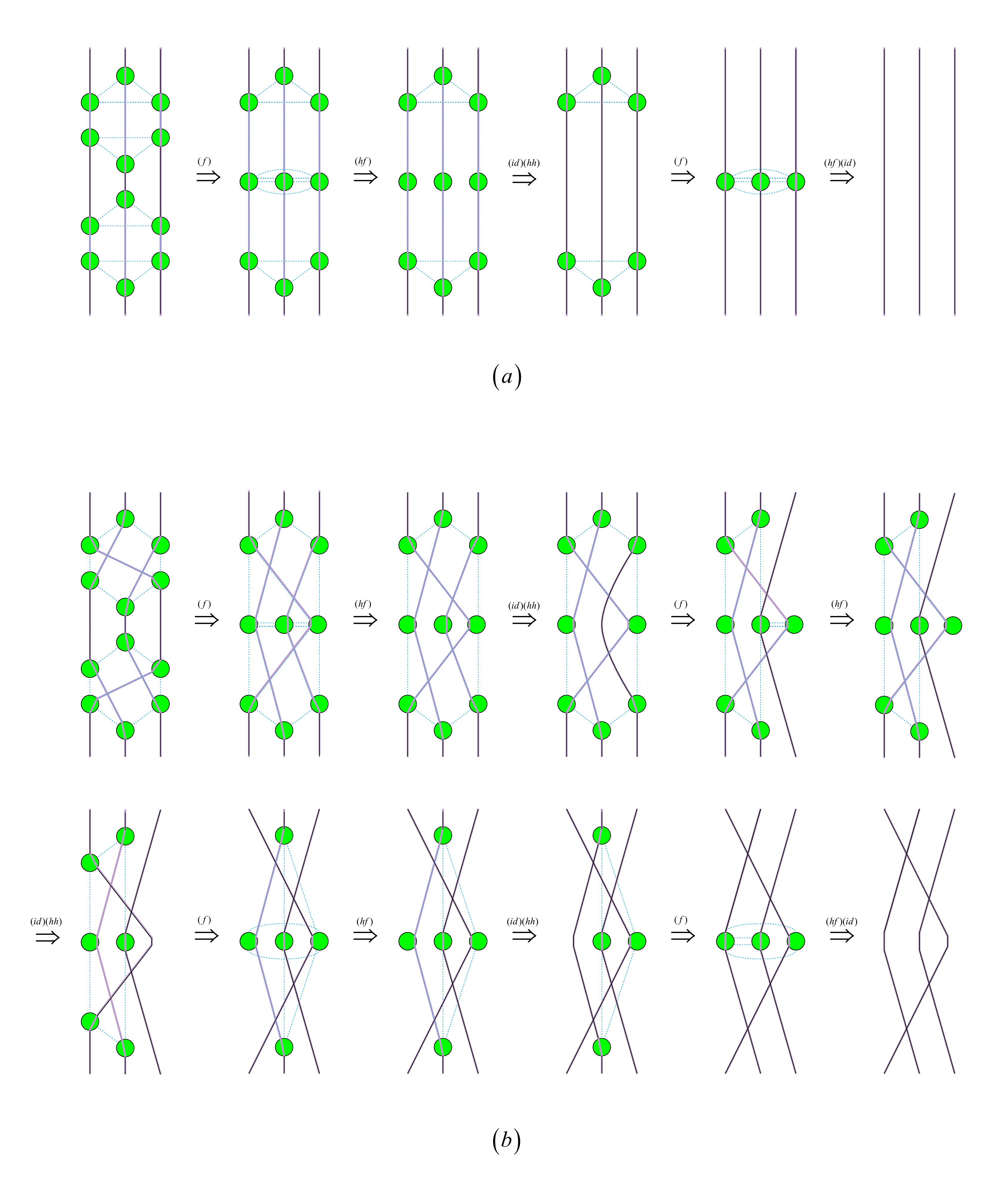}
\caption{ZX derivations of the perfect-tensor relation for two different choices of three contracted legs. The rewrite rules used at each step are indicated in the figure.} \label{fig-per}
\end{figure}

Now consider Figs.~\ref{fig-per}(a) and~\ref{fig-per}(b). They simply make explicit, at the level of string-diagram rewriting, what is treated as the perfect-tensor ``black box'' in the original HaPPY construction: each gives a proof of Eq.~\eqref{per}, with a different choice of three contracted legs.
The local rewrite rules invoked at each step are indicated directly in the figures. As these rules are applied successively, the initially complicated string diagram obtained by contracting two copies of the triangular-prism representation is gradually reduced: spiders fuse, pairs of blue lines cancel... In the end, the composite diagram reduces to three mutually tensor-decoupled bare wires, thereby giving an explicit ZX derivation of the perfect-tensor relation~\eqref{per}.

The preceding analysis is not tied to the particular triangular-prism ZX representation used above. Other ZX representations of the AME$(6,2)$ state can be analyzed in the same way, although the detailed rewrite pattern will of course depend on the chosen representation. Here we briefly introduce the standard procedure by which various kinds of ZX representations of AME states are obtained. Firstly, one can write the AME state in graph-state form~\cite{Helwig:2013qoq}. A graph state is specified by an ordinary graph $ G=(V,E), $ which should not itself be confused with the string diagrams used here. Each vertex represents a qubit prepared in the state $|+\rangle$, and after preparing $|+\rangle^{\otimes |V|}$, a CZ gate is applied along every edge of the graph.
Secondly, there is a standard translation from a graph state into a ZX diagram \cite{Duncan:2009nrf,Backens:2013hto}. Each vertex of $G$ is replaced by a $Z$-spider, each edge of $G$ by an edge carrying a Hadamard gate, and one Hilbert-space wire is extended from each spider. Applying this prescription to the triangular-prism graph-state representation of AME$(6,2)$ given in Ref.~\cite{Helwig:2013qoq} immediately yields the triangular-prism ZX representation shown in Fig.~\ref{fig-tpr}.
Therefore, the representation is not unique--since AME$(6,2)$ admits more than one graph-state representation, it correspondingly admits more than one ZX representation. For example, Ref.~\cite{Munne:2022ubc} employs another representative, referred to as the wheel graph-state representation, in the study of holographic quantum error-correcting codes. The relations among different graph-state representatives can be described by the theory of local-Clifford equivalence of graph states~\cite{Nest:2004khg}.

\subsection{QIF and trajectory nonuniqueness}\label{appc2}

Figs.~\ref{fig-per}(a) and~\ref{fig-per}(b) contain more information than merely a new proof of Eq.~\eqref{per}. Suppose that one of the two triangular-prism six-spider diagrams is interpreted as the state of a single HaPPY cell (i.e., a radius-one HaPPY disk), while the other is interpreted as a Clifford operation acting on three of its boundary degrees of freedom. The same diagrams may then be read as nontrivial quantum protocol diagrams.
Of course, strictly speaking, the six-spider diagram interpreted as a state should have all of its wires bent upward by the C-J correspondence, as in Fig.~\ref{fig-tpr}. Moreover, if the task is the transfer of quantum messages, the parties must append the standard teleportation after the Bell pairs have been exposed. Nevertheless, Figs.~\ref{fig-per}(a) and~\ref{fig-per}(b) already contain the essential string-diagram rewriting structure by which the protocol is eventually reduced to decoupled bare wires.
They therefore provide a particularly transparent example of how QIF trajectories are identified through compatible inheritance. The QIF trajectories are marked in purple in the figures. In the final, simplest representation, each purple trajectory is trivially supported on a decoupled bare wire. One may then trace it backward through the rewriting sequence.
At each step, a rewrite of the full morphism amounts locally to replacing some subdiagram $P$ by another subdiagram $P'$, while leaving the surrounding context $C[-]$ unchanged. Since $P$ and $P'$ represent the same morphism, their wire ports connecting to the external context are preserved. Thus, if an apparent through-path $\Gamma_k$ in $D_k$ enters and leaves $P'$ through ports $a$ and $b$, respectively, its compatible predecessor in $P$ must likewise connect the same two ports and, in addition, must not cross any tensor-product gap within $P$. This is the basic intuition behind the compatible-inheritance relation formalized in Ref.~\cite{Lin:2026hpd}. The purple trajectories in the initial diagrams of Figs.~\ref{fig-per}(a) and~\ref{fig-per}(b) are obtained precisely by tracing backward, step by step, from the bare wires in the terminal diagrams according to these inheritance rules.

\begin{figure}[t]
    \centering
    \includegraphics[width=\columnwidth]{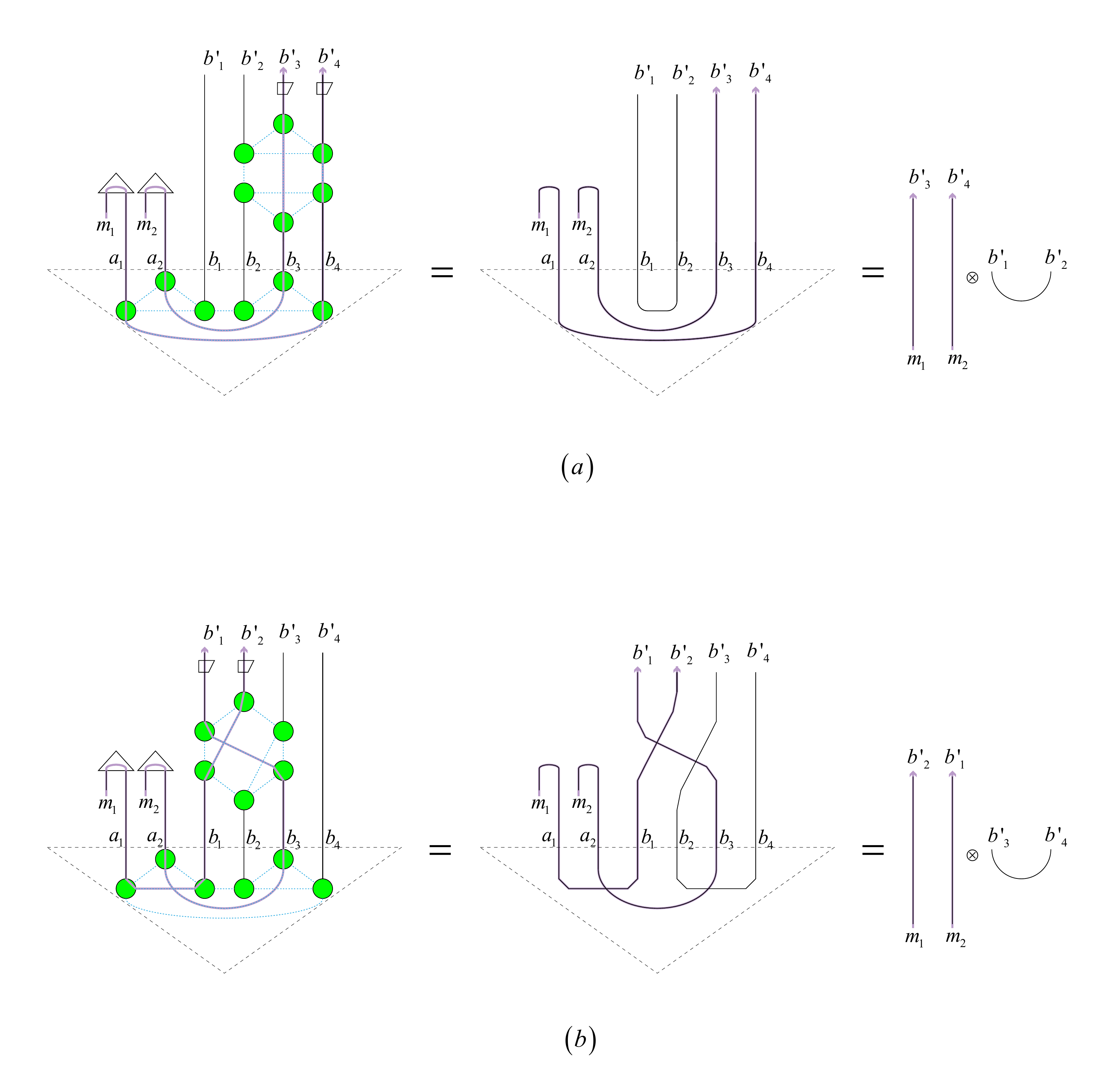}
\caption{Two QIF configurations for the same single-cell HaPPY state and the same two-party transfer task. Purple lines mark the corresponding trajectories in the resource-state sector.}\label{fig-sing}
\end{figure}

Let us make the physical content more explicit using the single-layer HaPPY disk shown in Fig.~\ref{fig-sing}.\footnote{In contrast with the usual conventions of the tensor-network literature, we draw all external legs upward so as to keep their string-diagram types explicit. Topological deformations of the drawing actually do not affect the argument here.} Divide the six boundary legs into
$
a_1,a_2\in A,
\qquad
b_1,b_2,b_3,b_4\in B.
$
For narrative convenience, one may think of these boundary degrees of freedom as ``experimenters'' belonging to two laboratories, $A$ and $B$. We ask how many message qubits laboratory $A$ can transfer to laboratory $B$.
Eq.~\eqref{per} already tells us that the answer is two~\cite{Pastawski:2015qua}. But we may ask a finer question: how can this transfer be characterized in terms of QIF configurations?
The rewrite chains of Figs.~\ref{fig-per}(a) and~\ref{fig-per}(b), beyond proving perfectness, immediately provide two answers, displayed in Figs.~\ref{fig-sing}(a) and~\ref{fig-sing}(b). In the first implementation, the parties associated with $b_2,b_3,b_4$ jointly apply a $B$-local Clifford operation represented by a triangular-prism ZX diagram whose ``height'' is vertically oriented, while $b_1$ and laboratory $A$ act trivially. In the second implementation, the parties associated with $b_1,b_2,b_3$ jointly apply a $B$-local operation represented by the same triangular-prism structure with a different, tilted choice of ``height'', while $b_4$ and laboratory $A$ again act trivially.
The resulting QIF trajectories are shown by the purple lines. If QIFs are taken to provide a concrete physical realization of bit threads, the two protocols therefore yield two distinct bit-thread configurations on the same single-cell HaPPY network, obtained simply by restricting the certified QIF trajectories to the tensor-network-state sector. In the first implementation, the two trajectories run
$
a_1\longrightarrow b_4,
\qquad
a_2\longrightarrow b_3.
$
Here the resource-state sector contains only one single bulk tensor vertex at the coarse-graining level. In the second implementation, they run
$
a_1\longrightarrow b_1,
\qquad
a_2\longrightarrow b_3.
$

This simple example already points to an important subtlety. The democratic isometry property of Eq.~\eqref{per} for every $k|6-k$ reading can easily tempt one to attribute more microscopic symmetry to an AME state than is actually present. An AME$(6,2)$ state need not possess all of the rotational and reflection symmetries suggested by drawing the cell as a regular hexagon. This becomes immediately apparent once one writes down a specific Dirac state vector or a concrete ZX realization, such as the triangular-prism representation used here, which visibly carries a preferred ``height'' direction.
Thus, once the HaPPY tensor ``black box'' is resolved microscopically, neither perfect hexagonal symmetry nor a unique microscopic realization should be expected. Different realizations may carry different detailed entanglement structures. QIFs probe precisely some of this finer structure that is easily hidden by equivalence at the level of entropy.
In particular, the perfectness relation~\eqref{per} alone is not sufficient to determine QIF or bit-thread trajectories. Such trajectory information depends on finer properties of the resource-state entanglement structure. A QIF trajectory--and hence a bit-thread trajectory when interpreted through the correspondence proposed here--is not a path that may simply be drawn by convention on the network. It must satisfy the nontrivial certification conditions of QIF, and therefore depends on whether the underlying microscopic entanglement structure actually provides the corresponding potential ``passage'' through the network.

\subsection{Geodesic-like bit-thread configurations}
\label{appc3}

\begin{figure}[t]
    \centering
    \includegraphics[width=\columnwidth]{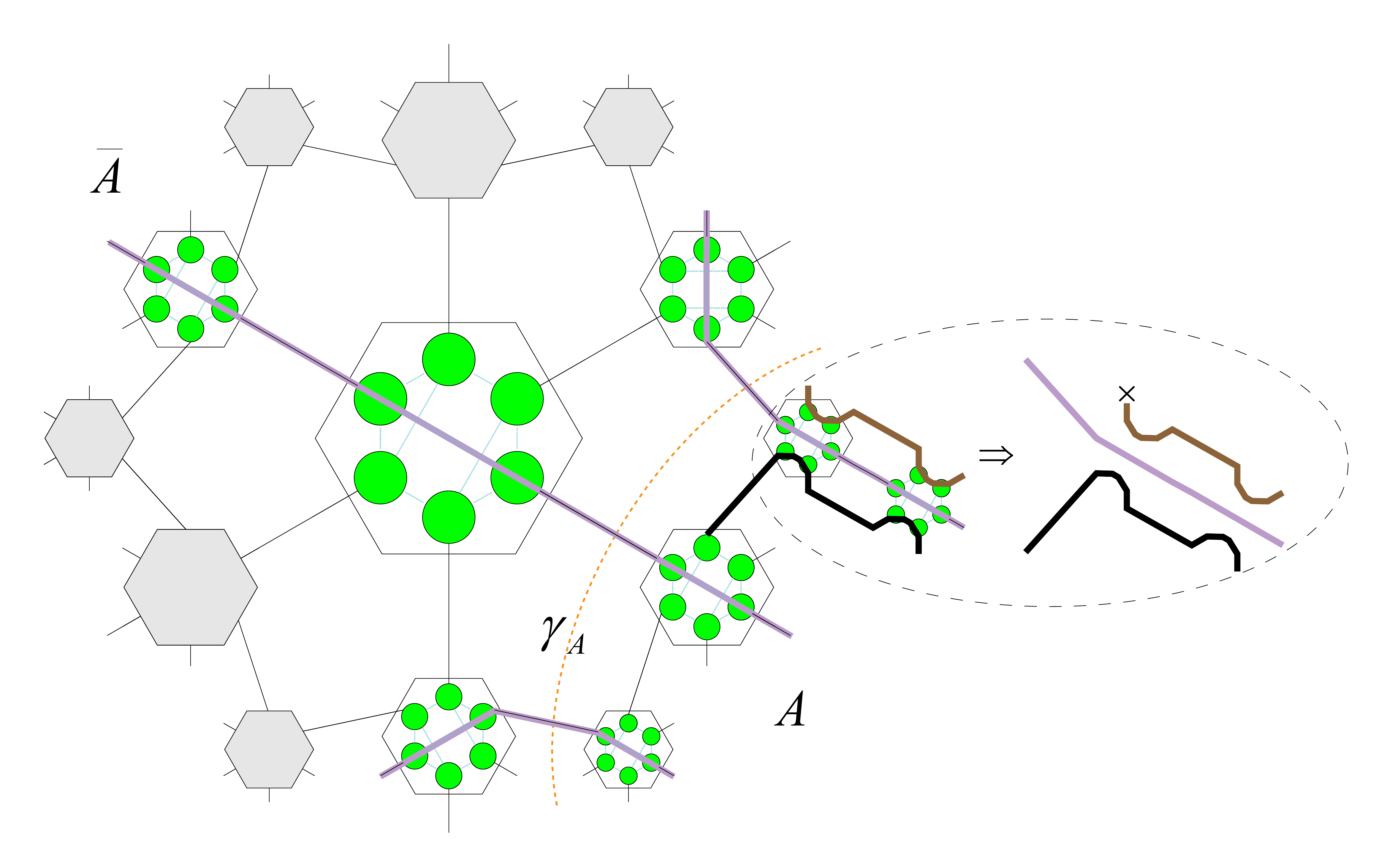}
\caption{A geodesic-like QIF configuration in a HaPPY network with suitably chosen microscopic ZX realizations. The inset illustrates one elementary peeling step.}\label{fig-dis}
\end{figure}

We now ask a further question. Can the familiar geodesic-like bit-thread configurations discussed in the literature~\cite{Agon:2018lwq} be realized in the HaPPY model as quantum information flows?
For suitable microscopic realizations, the answer is yes. Fig.~\ref{fig-dis} illustrates one such construction: a geodesic-like QIF configuration can be certified provided that the microscopic realizations of certain AME cells in the HaPPY network are chosen appropriately. Although the figure is drawn schematically as a simple two-layer disk,\footnote{For convenience in this subsection, we do not insist on drawing every boundary leg of the resource state strictly upward, nor do we keep track pictorially of every input/output orientation of the $3\to3$ unitary maps $T^\dagger$. This graphical shorthand should not cause ambiguity.} the construction expresses a more general idea.
More precisely, the purple through-paths drawn along the geodesic-like trajectories in Fig.~\ref{fig-dis} can indeed be certified as the resource-state-sector portions of QIFs generated by an admissible protocol. In other words, these paths can be compatibly inherited nontrivially throughout the string-diagram rewriting sequence until, in the terminal representation, they are supported on bare wires.

The reader will notice that we do not display a microscopic ZX realization for every hexagonal cell. Instead, we specify particular ZX-spider arrangements only for those cells traversed by the through-paths to be certified. In fact, such an arrangement also fixes a particular assignment of the physical legs. This choice is deliberate and is tailored to the protocol constructed here.
The construction of the protocol is straightforward. It closely parallels the iterative distillation procedure described in Ref.~\cite{Pastawski:2015qua}: one successively peels off outermost AME cells by applying appropriate $3\to3$ operations $T^\dagger$, until the entire resource state is reduced, by operations local to the $A$ and ${\bar A}$ sides, to $S(A)$ decoupled bare wires--three in the schematic example of Fig.~\ref{fig-dis}.
The inset of Fig.~\ref{fig-dis} shows the elementary peeling step. The two rewrite chains in Fig.~\ref{fig-per} may be used as ``macro rules'': for a given cell, one chooses a $3\to3$ operation $T^\dagger$ whose ZX realization is the vertical mirror of the cell realization, so that the corresponding network fragment reduces to three bare wires. If both ends of one such bare wire are free, as for the brown wires in the figure, it may be removed by a trivial local discard. If one end remains attached to the unreduced network, the wire is either used to support a QIF trajectory, marked in purple, or retained as a wire on which a later peeling step will act, marked in black.
Iterating this procedure reduces the network to precisely $S(A)$ parallel bare wires supporting the QIFs that have been compatibly inherited throughout the construction. This completes the certification.
The essential point is that, in order for a geodesic-like QIF trajectory to be certified, the microscopic ZX realization of each AME cell crossed by the proposed through-path must be arranged so that the through-path left behind when that cell is peeled off follows the desired geodesic-like route. One explicit way to achieve this is to orient the ``height'' of the triangular-prism ZX realization of each such AME cell along that route.

This immediately raises a natural question. Why should the appearance of a geodesic-like QIF seem to require a special microscopic realization of the HaPPY network? If the triangular-prism ``height'' were oriented differently in each cell, or if one replaced the cell by another concrete realization of the same AME state--for example, a wheel-type ZX representation~\cite{Helwig:2013qoq}--would a geodesic-like QIF still necessarily exist?
In general, it need not. But this is precisely the kind of microscopic distinction that the QIF analysis is meant to expose.
To see why, one should reconsider the original question. There is no a priori principle of quantum information theory requiring a quantum information flow to follow a geodesic. The motivation comes instead from the Headrick-Freedman bit-thread framework, in which geodesic-like optimal flows constitute one of the most natural and familiar classes of configurations~\cite{Agon:2018lwq}. Yet Headrick-Freedman threads arise originally as mathematical objects obtained from the convex-optimization dual of the RT formula. If these objects are to bear a genuine quantum-information interpretation, their microscopic physical meaning must be investigated rather than assumed. The present construction should be understood as one step in that direction.

This leaves at least two logically distinct possibilities.
First, it may be that only a subset of mathematically admissible Headrick--Freedman thread configurations admit a QIF realization and hence the particular operational meaning proposed here. Other configurations may remain purely mathematical convex-optimization flows, or may require a different physical interpretation.

Second, one may entertain a more optimistic--although considerably stronger--possibility: perhaps every thread configuration admitted by the Headrick-Freedman framework reflects a QIF configuration that the entanglement structure of a genuine holographic state ought, in some appropriate sense, to support. If this were true, bit-thread configurations could be used in the reverse direction, as data constraining the microscopic entanglement structure of holographic states.
For the HaPPY model in particular, this possibility suggests an interesting lesson. A fixed microscopic realization of the cell tensors need not reproduce every desired QIF configuration. A more faithful holographic model might therefore require, for example, a quantum superposition, an averaging procedure, or an equivalence class of microscopic configurations. Alternatively, there may exist some as-yet-unidentified gauge principle under which different microscopic presentations are merely different gauge appearances of the same underlying physical structure, with the full family of bit-thread configurations recovered only at that more invariant level.

\section{QIF in holographic stabilizer models}\label{appd}

\subsection{Bell-block identification and local Clifford standardization}
\label{appd1}

For holographic tensor-network states belonging to the stabilizer family\cite{Pastawski:2015qua, Hayden:2016cfa, Nezami:2016zni}, Ref.~\cite{Fattal:2004frh} has shown that, for any pure bipartite stabilizer state
$
|\Psi\rangle_{AB},
$
there is a systematic procedure for identifying the ``Bell blocks'' that cross the bipartition and, from them, constructing local Clifford operations $U_A$ and $U_B$ on the two sides such that the original state is brought into the canonical form
\begin{equation}
(U_A\otimes U_B)|\Psi\rangle_{AB}
=
|\Phi^+\rangle_{A'B'}^{\otimes k}
\otimes
|\alpha\rangle_{A_{\rm junk}}
\otimes
|\beta\rangle_{B_{\rm junk}} .
\label{stab-canonical}
\end{equation}
Here
$
|\Phi^+\rangle=\frac{|00\rangle+|11\rangle}{\sqrt2}
$
is a Bell state, $|\alpha\rangle$ and $|\beta\rangle$ are local stabilizer junk states supported entirely within $A$ and $B$, respectively, and
$
k=\frac{S(A)}{\log 2}
$
is the bipartite entanglement entropy measured in qubits, equivalently the number of Bell pairs that can be extracted.
The basic idea of Ref.~\cite{Fattal:2004frh} is that, for a bipartite stabilizer state, the stabilizer generators can be organized into local parts, supported only on $A$ or only on $B$, and a nonlocal part crossing the bipartition. The nonlocal generators can be chosen in canonical pairs; in the terminology adopted here, these are precisely the \emph{Bell blocks}. Once such blocks have been identified, the remaining task for our purposes is to find local Clifford operations on the two sides that standardize them into Bell pairs. We now describe these two technical steps more explicitly: identifying Bell blocks from the stabilizer generators, and constructing local Clifford operations that bring the identified blocks into standard Bell-pair form.

We begin with the basic picture of a Bell block. The standard stabilizers of
$
|\Phi^+\rangle_{ab}
=
\frac{|00\rangle+|11\rangle}{\sqrt2}
$
may be taken to be
$X_aX_b$, $Z_aZ_b.$
If $a\in A$ and $b\in B$, these two stabilizers have an important structure. On the $A$ side, $X_a$ and $Z_a$ anticommute; on the $B$ side, $X_b$ and $Z_b$ also anticommute. Globally, however, $X_aX_b$ and $Z_aZ_b$ commute, since the two anticommutation signs contributed by the two sides cancel. Thus the stabilizer structure of a Bell pair may be summarized succinctly as
$
\text{locally anticommuting, globally commuting}.
$
For a general bipartite stabilizer state, a Bell block is a nonstandard realization of precisely this structure. Namely, there is a pair of nonlocal stabilizer generators $g,\bar g$ that commute globally, while their restrictions to $A$ anticommute and their restrictions to $B$ also anticommute. Such a pair of generators is the abstract stabilizer precursor of a Bell pair.
More explicitly, for a stabilizer group
$
\mathcal S=\langle g_1,\ldots,g_n\rangle
$
and a bipartition $A|B$, one first organize the stabilizer group into three sectors, $
\mathcal S_A$, $
\mathcal S_B$, $
\mathcal S_{AB}.
$
Here $\mathcal S_A$ is supported only on $A$ and eventually contributes to the $A$-side local junk; $\mathcal S_B$ is supported only on $B$ and eventually contributes to the $B$-side local junk; and $\mathcal S_{AB}$ is the genuinely nonlocal part crossing the bipartition, whose generators can be organized into Bell blocks and subsequently brought into standard Bell-pair form by local Clifford operations.
For a generic choice of stabilizer generators, the Bell-block structure need not be manifest. Some generators may still contain components associated with Bell blocks that have already been selected. By systematically rechoosing the stabilizer generating basis, these mixed components can be eliminated and the nonlocal anticommuting pairs isolated~\cite{Fattal:2004frh}. In the binary symplectic representation of Pauli operators, this procedure may naturally be viewed as a Gram-Schmidt -like reduction with respect to the Pauli symplectic form.
\footnote{Up to phase, an $n$-qubit Pauli string can be represented by a binary vector
$(x|z)\in\mathbb F_2^{2n},$ with its commutation and anticommutation relations encoded by the corresponding binary symplectic form~\cite{Dehaene:2003thc}. This symplectic structure therefore provides a systematic way to search for and isolate the nonlocal anticommuting pairs.}

Once the Bell blocks have been identified, each block determines a local Pauli pair on either side,
$P_i^A$, $ Q_i^A$; $P_i^B$, $ Q_i^B$,
with the same symplectic commutation relations as the standard pair $X_i,Z_i$. The next task is therefore to construct local Clifford transformations implementing
\begin{equation}
P_i^A,Q_i^A
\longmapsto
X_{a_i},Z_{a_i},
\qquad
P_i^B,Q_i^B
\longmapsto
X_{b_i},Z_{b_i}.
\label{pauli-standardization}
\end{equation}
For the GHZ example below, this standardization can be recognized immediately as a single CNOT gate. In the general case, however, $P_i^A,Q_i^A$ and $P_i^B,Q_i^B$ may be long Pauli strings supported on several qubits. Their Clifford standardization can then be carried out systematically by a procedure closely analogous to ordinary Gaussian elimination~\cite{Dehaene:2003thc,Aaronson:2004xuh}. Schematically, one first chooses a pivot qubit and uses $H$ and $S$ gates to adjust the Pauli type on the pivot. Local Clifford gates such as CNOT, CZ, and SWAP are then used to eliminate unwanted Pauli support on the remaining qubits. In this way, a nonstandard Pauli pair can be reduced to the canonical pair $X,Z$ on a single qubit.
The procedure closely parallels the elimination of entries in ordinary Gaussian elimination using a pivot column. The essential difference is that the allowed operations here must preserve the Pauli symplectic commutation relations, and are therefore implemented by Clifford gates. Repeating the procedure for the remaining Bell blocks completes the standardization. By recording these elementary Clifford gates step by step, the resulting $U_A$ and $U_B$ are obtained not merely as abstract existence statements, but as explicit gate-level circuits.

\subsection{ZX representation of Clifford quantum circuits}\label{appd2}

\begin{figure}[t]
    \centering
    \includegraphics[width=\columnwidth]{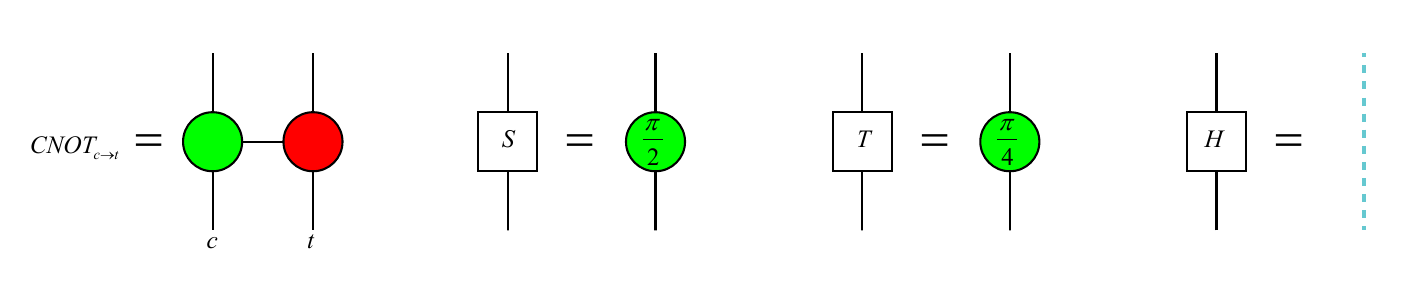}
    \caption{ The ZX representation of the CNOT, $S$, $T$ and $H$ gates.}
    \label{fig-cnot}
\end{figure}

In quantum information theory, one often says that a small collection of elementary gates is sufficient to construct very general quantum circuits~\cite{Nielsen:2012yss}. More precisely, two related notions of universality are commonly involved.
First, if arbitrary single-qubit unitaries are allowed together with CNOT gates between arbitrary pairs of qubits, where
\begin{equation}
\mathrm{CNOT}_{c\to t}
=
|0\rangle\langle0|_c\otimes I_t
+
|1\rangle\langle1|_c\otimes X_t ,
\label{cnot-def}
\end{equation}
then an arbitrary $n$-qubit unitary transformation can in principle be constructed exactly. This is universality with a continuous gate set. For this reason, it is useful to recall in particular the ZX representation of the CNOT gate~\cite{vandeWetering:2020giq,Kissinger:2024pqs}. With the convention adopted throughout this work that string diagrams are read from bottom to top, the standard ZX representation of CNOT is shown in Fig.~\ref{fig-cnot}: a zero-phase green spider is placed on the control wire, a zero-phase red spider on the target wire, and the two spiders are connected.

A second notion is approximate universality using a finite gate set. A standard example is
$
\{H,S,T,\mathrm{CNOT}\},
$
often referred to as Clifford+$T$, with
\begin{equation}
H=\frac{1}{\sqrt2}
\begin{pmatrix}
1&1\\
1&-1
\end{pmatrix},
\qquad
S=
\begin{pmatrix}
1&0\\
0&i
\end{pmatrix},
\qquad
T=
\begin{pmatrix}
1&0\\
0&e^{i\pi/4}
\end{pmatrix}.
\end{equation}
Although a finite gate set cannot represent every continuous unitary matrix exactly, it can approximate an arbitrary target unitary to arbitrary accuracy. That is, for any target unitary $U$ and any $\epsilon>0$, one can construct a circuit $V$ from these elementary gates such that
\begin{equation}
\|U-V\|<\epsilon .
\end{equation}
The ZX representations of the $S$ and $T$ gates are also shown in Fig.~\ref{fig-cnot}; they are two-legged green spiders with phases $\pi/2$ and $\pi/4$, respectively.

For the purposes of this work, we restrict attention to the Clifford fragment, i.e., circuits generated without the $T$ gate and the corresponding stabilizer states.
This sector is sufficiently nontrivial to exhibit the structures of interest here, while remaining systematically tractable, and it plays an important role in holographic tensor-network models.

\subsection{GHZ as a minimal stabilizer prototype: constructing ZX diagrams from Bell blocks}\label{appd3}

We now use the GHZ state as a prototype to illustrate the minimal version of the full mechanism described above. Let
$
|\mathrm{GHZ}\rangle_{123}
=
\frac{|000\rangle+|111\rangle}{\sqrt{2}},
$
and consider the bipartition
$
A=\{1\}$, $
B=\{2,3\}.
$
Choose the following generating set for the GHZ stabilizer group:
$
g_1=X_1X_2X_3$, $
g_2=Z_1Z_2$, $
g_3=Z_2Z_3.
$
We now show how the local part and the nonlocal Bell block can be read directly from this stabilizer structure. The first observation is that
$
g_3=Z_2Z_3
$
acts only on the $B$ side. It therefore belongs to the $B$-local stabilizer sector,
$
\mathcal S_B=\langle Z_2Z_3\rangle,
$
and will eventually contribute to the $B$-side local junk. On the other hand, the $A$ side contains only qubit $1$, and none of the three generators contains a nontrivial stabilizer supported solely on that qubit. Hence
$
\mathcal S_A=\{I\}.
$
This leaves the two generators
$
g_1=X_1X_2X_3$, $
g_2=Z_1Z_2,
$
as the natural candidates for the nonlocal Bell block. We therefore inspect their restrictions to the two sides of the bipartition. On $A$,
$
g_1|_A=X_1$, $
g_2|_A=Z_1,
$
so the two restrictions anticommute. On $B$,
$
g_1|_B=X_2X_3$, $
g_2|_B=Z_2,
$
and these again anticommute, since $X_2$ anticommutes with $Z_2$. Globally, however, $g_1$ and $g_2$ commute: the $A$ and $B$ sides each contribute one anticommutation, and the two minus signs cancel. This is exactly the structure we were looking for. Thus
$
(g_1,g_2)
=
(X_1X_2X_3,Z_1Z_2)
$
constitutes a Bell block.
The GHZ stabilizer group can therefore be read as
\begin{equation}
\mathcal S
=
\underbrace{\langle Z_2Z_3\rangle}_{B\text{-side local junk}}
\oplus
\underbrace{\langle X_1X_2X_3,Z_1Z_2\rangle}_{\text{one Bell block}}.
\label{ghz-block1}
\end{equation}
This is the minimal realization of the canonical decomposition of Ref.~\cite{Fattal:2004frh}. It tells us that, across the $1|23$ bipartition, the GHZ state contains one extractable Bell pair together with a $B$-local junk degree of freedom. In essence, the procedure amounts to rereading the stabilizer group in a way that makes its entanglement structure manifest.

Once the Bell blocks have been identified, one can in principle find the required local Clifford operations on the two sides that bring them into standard Bell-pair form. In the present GHZ example, we have identified the Bell block
$
(X_1X_2X_3,\;Z_1Z_2).
$
On the other hand, a standard Bell pair $|\Phi^+\rangle_{12}$ is stabilized by
$
X_1X_2$, $ Z_1Z_2.
$
Thus all that remains is to standardize the $B$-side Pauli pair
$
X_2X_3,\;Z_2
\quad\longmapsto\quad
X_2,\;Z_2.
$
It can be proved immediately that this is simply achieved by the $B$-local Clifford gate
$
\mathrm{CNOT}_{2\to3}.
$
Indeed, its action by Pauli conjugation includes
$
X_2X_3\longmapsto X_2$, $
Z_2\longmapsto Z_2.
$
Consequently,
$
X_1X_2X_3\longmapsto X_1X_2$, $
Z_1Z_2\longmapsto Z_1Z_2,
$
while the local junk generator is simultaneously mapped as
$
Z_2Z_3\longmapsto Z_3.
$
The final stabilizer group is therefore generated by
$
X_1X_2$, $ Z_1Z_2$, $ Z_3,
$
which stabilizes
$
|\Phi^+\rangle_{12}\otimes|0\rangle_3.
$
Hence
\begin{equation}
\mathrm{CNOT}_{2\to3}|\mathrm{GHZ}\rangle_{123}
=
|\Phi^+\rangle_{12}\otimes|0\rangle_3.
\label{ghz-decoder1}
\end{equation}

Thus, for this simple GHZ example, we have completed the construction: following the canonical-decomposition strategy of Ref.~\cite{Fattal:2004frh}, we have obtained an explicit LOCC protocol that transforms the resource state into a Bell pair together with decoupled junk.
\begin{figure}[t]
    \centering
    \includegraphics[width=\columnwidth]{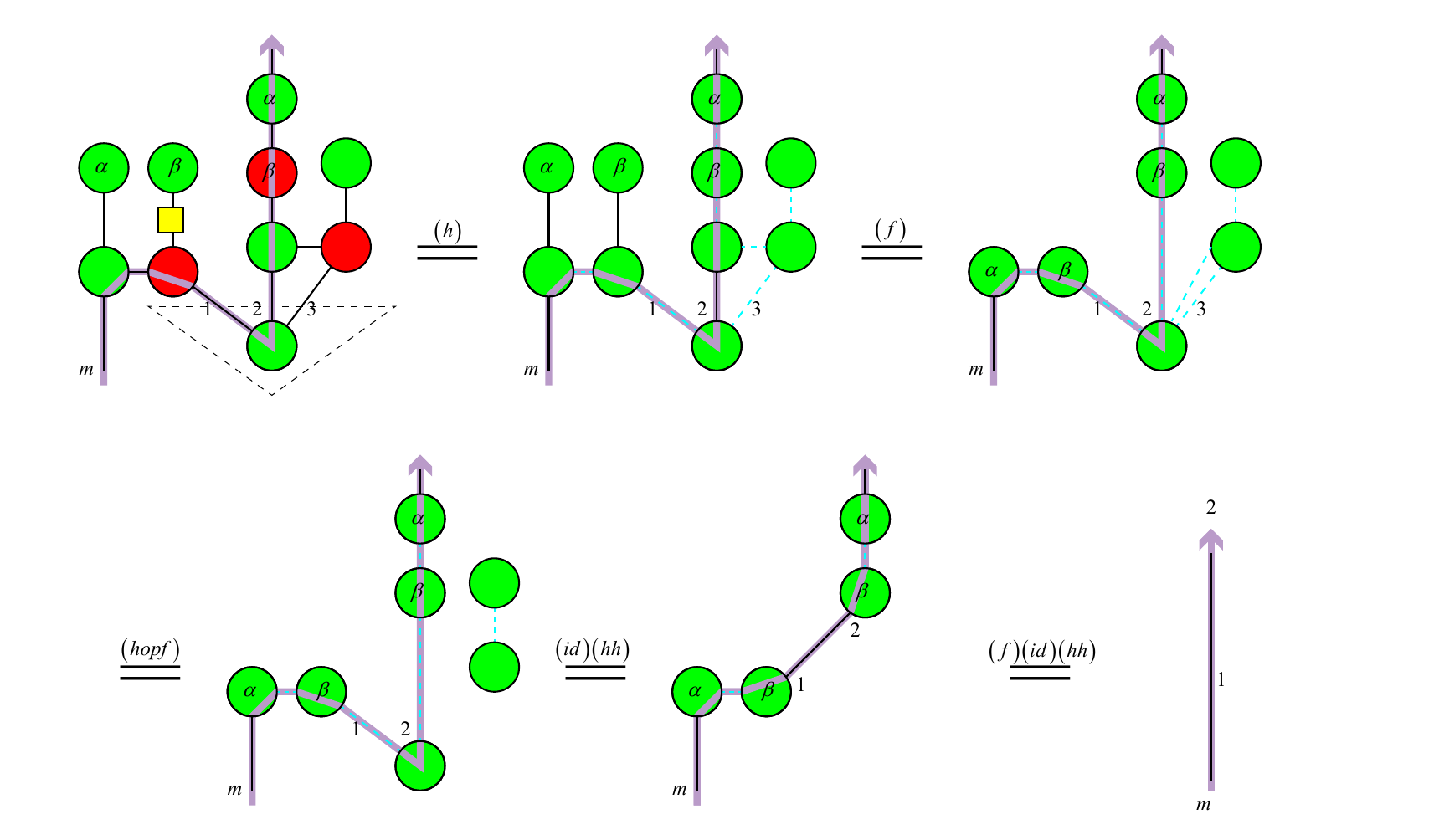}
\caption{ZX certification of the QIF associated with the first Bell-block construction. Its trajectory through the GHZ resource state runs from leg $1$ to leg $2$.}
    \label{fig-ghz1}
\end{figure}

We next translate the CNOT decoder into its ZX-spider representation, as described in Appendix~\ref{appd2}, and add the teleportation components required to turn the above Bell-pair extraction scheme into a complete protocol for transferring the message qubit from $A$ to $B$. We also append a trivial zero-phase spider as an effect, solely to remove the final decoupled junk. This gives the complete initial string diagram shown in Fig.~\ref{fig-ghz1}.
Figure~\ref{fig-ghz1} further displays an explicit sequence of ZX rewrites, which indeed certifies a QIF. In the present construction, the certified trajectory starts at the message input $m_1$, enters the resource-state sector through boundary leg $1$, and exits through boundary leg $2$.

\begin{figure}[t]
    \centering
    \includegraphics[width=\columnwidth]{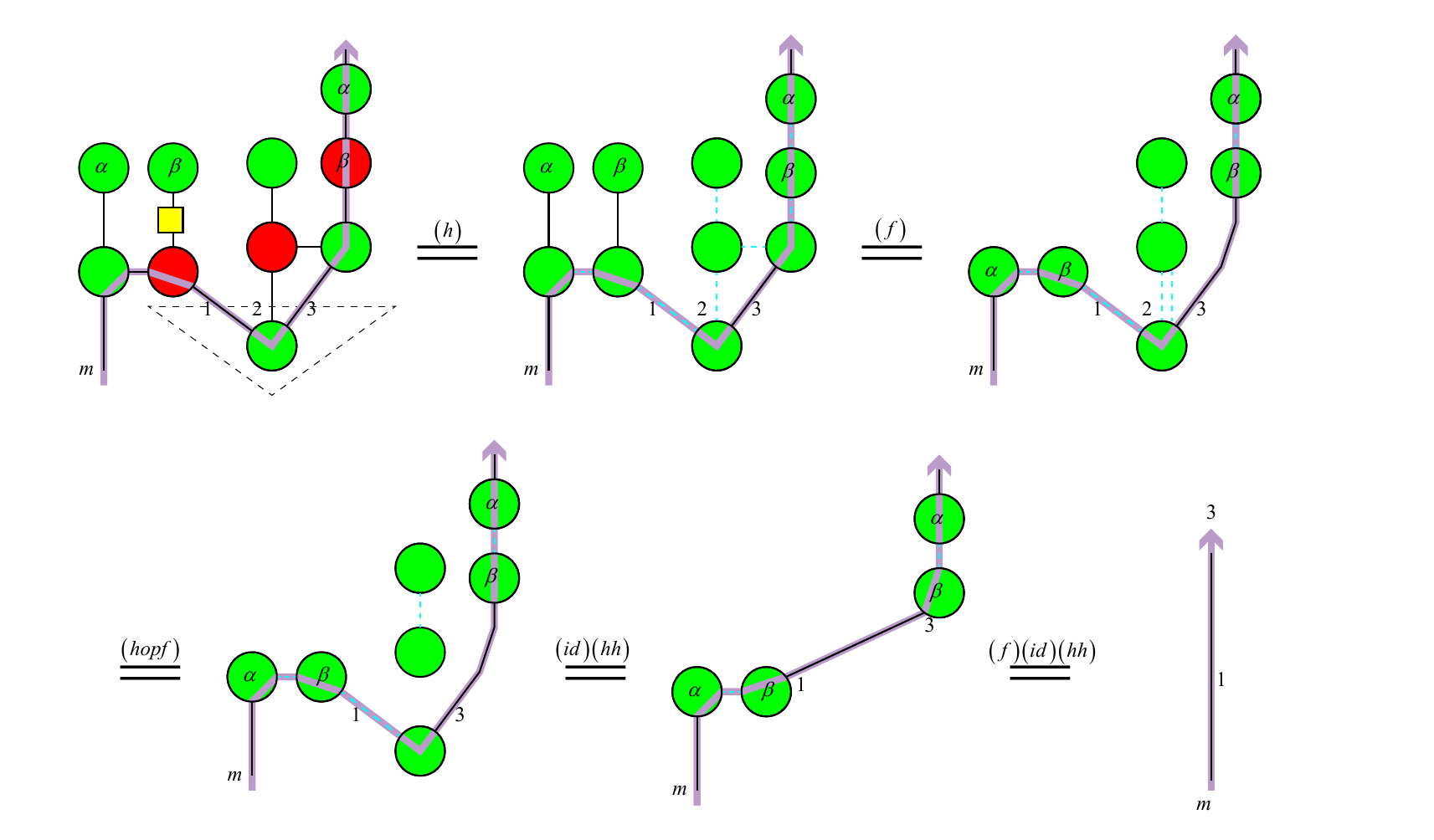}
\caption{ZX certification of the QIF associated with the second Bell-block construction. Its trajectory through the GHZ resource state runs from leg $1$ to leg $3$.}
    \label{fig-ghz2}
\end{figure}

We now point out that the same GHZ state admits another Bell-block reading. Define
\begin{equation}
g_2'
:=
g_2g_3
=
Z_1Z_3 .
\label{g2prime}
\end{equation}
We may then choose
$
g_1$, $ g_2'$, $ g_3
$
as an alternative generating set for the same stabilizer group. Importantly, nothing about the GHZ state itself has changed; we have only changed the way in which its stabilizer structure is read.
Consider now the new nonlocal pair
$
(g_1,g_2')
=
(X_1X_2X_3,\;Z_1Z_3).
$
On the $A$ side,
$
g_1|_A=X_1$, $
g_2'|_A=Z_1,
$
so the two restrictions anticommute. On the $B$ side,
$
g_1|_B=X_2X_3$, $
g_2'|_B=Z_3,
$
and these again anticommute. Hence $(g_1,g_2')$ constitutes another valid Bell block.
This time, the required $B$-local Clifford operation is
\begin{equation}
U_B^{(2)}
=
\mathrm{CNOT}_{3\to2}.
\end{equation}
Its Pauli conjugation action gives
$
X_2X_3\longmapsto X_3$, $
Z_3\longmapsto Z_3.
$
Consequently,
$
g_1\longmapsto X_1X_3,
\qquad
g_2'\longmapsto Z_1Z_3,
$
while at the same time
$
g_3=Z_2Z_3\longmapsto Z_2.
$
We therefore obtain
\begin{equation}
\mathrm{CNOT}_{3\to2}
|\mathrm{GHZ}\rangle_{123}
=
|\Phi^+\rangle_{13}\otimes|0\rangle_2 .
\label{ghz-decoder2}
\end{equation}
Proceeding exactly as before then gives the ZX string diagram shown in Fig.~\ref{fig-ghz2}. In this second construction, the certified QIF again starts at $m_1$ and enters the resource-state sector through boundary leg $1$, but now exits through boundary leg $3$. We have thus obtained a second implementation of the same task--transferring one message qubit from $A$ to $B$---which certifies a different QIF trajectory and, correspondingly, a different bit-thread trajectory in the resource-state sector.

Several remarks are in order. First, we do not claim that, for a bipartite stabilizer state, every protocol that transfers the maximal number of message qubits from $A$ to $B$ and certifies the corresponding QIFs must follow the systematic canonical-decomposition procedure described in this section. This is clearly not the case. Recall, for example, the two protocols constructed in Appendix~\ref{appb} for the same GHZ resource state, whose ZX realizations are given in Ref.~\cite{Lin:2026hpd}. These protocols are manifestly different from the two constructions obtained here through Bell-block identification followed by a CNOT decoder. The canonical-decomposition method developed in this section should therefore not be understood as an exhaustive classification of all possible QIF witnesses. Rather, it provides a systematic and constructive sufficient route, one that can be explicitly resolved into elementary gates and subsequently translated into ZX string diagrams.
On the other hand, the two protocols of Appendix~\ref{appb} and the two constructions presented here nevertheless give equally valid bit-thread trajectories in the resource-state sector--one may be viewed as running from $a_1$ to $b_1$, and the other from $a_1$ to $b_2$--while each complete protocol provides its own QIF witness. This distinction is worth emphasizing: a complete QIF witness and a bit-thread configuration in the resource-state sector are objects at different levels. The former depends on the full protocol, including its specific operations, whereas the latter is the projection of the QIF onto the resource-state sector. Consequently, the same bit-thread configuration may be realized by more than one concrete protocol.

\section{ Trajectory correspondence between QIFs and bit threads }\label{appe}

\subsection{Microscopic derivation of the trajectory correspondence}\label{appe1}

This section provide the details underlying the trajectory-level correspondence between quantum information flows and bit threads described in the main text.
A tensor network is usually, in a certain sense, a coarse-grained skeleton
$
\mathcal{G}=(V,E,c),
$
where $V$ is the set of macroscopic tensor vertices, $E$ is the set of macroscopic bonds, and $c_e\in\mathbb N$ is the qubit capacity of a bond $e$. Assume that the Hilbert space associated with each macroscopic bond $e$ can be refined into elementary qubit degrees of freedom as
$
\mathcal H_e\simeq (\mathbb C^2)^{\otimes c_e}.
$
Correspondingly,
\begin{equation}
B_e=\{e_{e,1},\ldots,e_{e,c_e}\},
\qquad
|B_e|=c_e,
\end{equation}
denotes the microscopic wire bundle associated with the macroscopic bond $e$. Each macroscopic tensor $v\in V$ is then expanded into a local process subdiagram $R_v$ connecting these wires. The same decomposition is applied to the operation tensors $O_A$ and $O_{\bar A}$ on the two sides. In this way, both the tensor network and the entire entanglement network are represented, at the technical level, by a semantics-preserving microscopic string diagram $\widetilde{\mathcal{G}}$, whose set of microscopic wires is denoted by $\widetilde E$ and whose set of microscopic nodes is denoted by $\widetilde V$. The existence of such an elementary string-diagram decomposition is, of course, an implicit assumption. In our work, however, the ZX string-diagram language provides an explicit realization of it for holographic stabilizer models.

Consider a deterministically successful quantum protocol. For each measurement branch, let the protocol-network string diagram be denoted by $D_0$, with the branch label temporarily suppressed, and write a semantics-preserving string-diagram rewriting sequence as
\begin{equation}
D_0\Rightarrow D_1\Rightarrow\cdots\Rightarrow D_n .
\label{rew}
\end{equation}
To say that the protocol contains $k$ quantum information flows means that there exists such a rewriting sequence for which the final diagram takes the simple form
\begin{equation}
D_n\simeq
\left(
\bigotimes_{i=1}^{k} I_{m_i\rightarrow b_i}
\right)
\otimes R .
\label{fin}
\end{equation}
As illustrated in Fig.~\ref{fig-main}(b), $I_{m_i\rightarrow b_i}$ denotes a bare identity wire from an incoming wire $m_i$ to an outgoing wire $b_i$, while $R$ denotes the remaining junk diagram decoupled from these bare wires.
For the protocols involving only simple bipartite entanglement considered in Coecke's original discussions~\cite{Coecke:2004sxv,Coecke:2005bin}, the rewriting sequence~\eqref{rew} is usually rather short. As a result, the QIF trajectories in $D_0$ can be identified almost directly from the terminal form~\eqref{fin}, and can be described by assigning local traversal rules to the elementary component boxes. However, general multipartite entanglement and sufficiently complicated rewriting sequences both call for a more careful formalization. This was developed in Ref.\cite{Lin:2026hpd} as follows. 
First, in a specified string-diagram representation $D_i$, a path is said to be an apparent through-path if it does not cross any visible tensor-product gap in $D_i$; that is, it never passes directly between subdiagrams that occur as distinct tensor factors rather than being diagrammatically connected.
Second, one must formalize how an apparent through-path is compatibly inherited when the same morphism is represented by the next diagram $D_{i+1}$. ~\cite{Lin:2026hpd} characterizes this by a relation associated with the rewrite
$
\rho_i:D_i\Rightarrow D_{i+1},
$
such that
$
(\Gamma_i,\Gamma_{i+1})\in\mathcal T_{\rho_i}.
$
Then, if an apparent through-path $\Gamma_0$ in the initial diagram $D_0$ admits, through $\mathcal T_{\rho_0}$, a corresponding apparent through-path representative in $D_1$, and this procedure can be continued through every rewriting step until its representative in $D_n$ is precisely the trivial path on the bare identity wire $I_{m_i\rightarrow b_i}$, we call $\Gamma_0$ a genuine through-path.
A physical protocol, of course, generally contains measurement branches. Hence, if the same branch-independent initial path $\Gamma_0$ is a genuine through-path for every fixed set of measurement outcomes, and in every branch is ultimately inherited onto a decoupled bare identity wire $I_{m_i\rightarrow b_i}$, then it is certified as a quantum-information-flow trajectory on the entanglement-network string diagram $D_0$.

In the following analysis, let a set of QIFs certified by~\eqref{rew} and~\eqref{fin} be $\mathcal F=\{\Gamma_r\}_{r=1}^{k}.$
We fix a chosen compatible-inheritance scheme along the rewriting sequence~\eqref{rew}. For the analysis below, we restrict attention to the portions of the trajectories $\Gamma_r$ that lie within the holographic resource-state sector, yet denote these portions by $\Gamma_r$ as well.
Note that a QIF trajectory, being a line-like through-going structure in a string diagram, need not possess an intrinsic orientation. To compare it with the oriented formulation of a discrete flow, we adopt the following convention: every certified trajectory $\Gamma_r$ is oriented from the boundary region $A$ toward its complement $\bar A$. Under the quantum-communication interpretation, this orientation corresponds to a message being sent from the $A$ side and reconstructed on the $\bar A$ side.~\footnote{ The conservation and capacity arguments below, however, require only this orientation convention and do not depend on that operational interpretation.}
After orienting every certified trajectory from $A$ to $\bar A$, we independently choosing a reference orientation for each microscopic wire $\tilde e$. Then we compare the direction in which the trajectories traverse each wire with its reference orientation. In this way, the fixed flow family $\mathcal F$ induces the following labels on the microscopic wire set:\begin{equation}
j_{\tilde e}
=
\begin{cases}
+1,
& \substack{
\tilde e\in\Gamma_r \text{ for some } r,\\
\Gamma_r \text{ traverses } \tilde e \text{ along its reference orientation},
}
\\[2mm]
-1,
& \substack{
\tilde e\in\Gamma_r \text{ for some } r,\\
\Gamma_r \text{ traverses } \tilde e \text{ against its reference orientation},
}
\\[2mm]
0,
& \tilde e \text{ is not traversed by the flow family}.
\end{cases}
\end{equation}
Thus
\begin{equation}
j_{\tilde e}\in\{-1,0,+1\}.
\end{equation}
This is simply a line-by-line registration of the fixed flow family on the microscopic string diagram.\footnote{If the conventional orientations of all trajectories are reversed simultaneously, then $j_{\tilde e}$ and the macroscopic flows $f_e$ introduced below both change sign, whereas the line-occupation numbers $n_e$, the divergenceless condition, and the capacity constraints remain unchanged. The trajectory correspondence therefore does not depend on choosing $A\rightarrow\bar A$ rather than $\bar A\rightarrow A$; the choice merely fixes the sign convention for the flow direction and boundary flux.}

Consider now an internal microscopic process node
$
a\in\widetilde V_v
$
within the tensor network state sector. To keep track of whether a microscopic wire is oriented into or out of $a$, define the signed incidence coefficient\begin{equation}
\epsilon_{a\tilde e}
=
\begin{cases}
+1, & \tilde e \text{ points away from } a,\\
-1, & \tilde e \text{ points toward } a,\\
0,  & \tilde e \text{ is not incident on } a.
\end{cases}
\end{equation}
If a trajectory $\Gamma_r$ does not pass through $a$, its net contribution at $a$ is clearly zero. If $\Gamma_r$ does pass through $a$, then because it is an apparent through-path in the current microscopic string diagram, it must enter along one wire incident on $a$ and leave along another. It can neither terminate at an internal process node nor cross a tensor-product gap exposed by the string diagram and jump to a disconnected process fragment. Hence every trajectory passing through $a$ contributes one incoming and one outgoing unit. Summing over all trajectories gives
\begin{equation}
\sum_{\tilde e\ni a}
\epsilon_{a\tilde e}j_{\tilde e}=0 ,
\label{mic1}
\end{equation}
for every internal microscopic node $a$.
We next derive the divergenceless property at the level of the macroscopic skeleton. For each macroscopic bond $e$, choose a reference orientation and orient all microscopic wires in the bundle $B_e$ consistently with it. Define the macroscopic net flow by
\begin{equation}
f_e:=
\sum_{\tilde e\in B_e}j_{\tilde e}.
\end{equation}
Now fix an macroscopic vertex $v$ in the tensor network and sum Eq.~\eqref{mic1} over all internal nodes in its microscopic realization $R_v$:
\begin{equation}
\sum_{a\in\widetilde V_v}
\sum_{\tilde e\ni a}
\epsilon_{a\tilde e}j_{\tilde e}=0 .
\end{equation}
Every internal wire $\tilde e\in\widetilde E_v^{\,\mathrm{int}}$ occurs exactly twice in this sum, once at each endpoint. The corresponding signed incidence coefficients have opposite signs, and the two contributions therefore cancel. 
After this cancellation, only microscopic wires crossing the boundary of $R_v$ remain. These are microscopic wires belonging to the bundles $B_e$ associated with the macroscopic bonds $e$ incident on $v$. Grouping the remaining terms by bond then gives
\begin{equation}
\sum_{e\ni v}\epsilon_{ve}f_e=0 ,
\label{mac1}
\end{equation}
where
\begin{equation}
\epsilon_{ve}
=
\begin{cases}
+1, & e \text{ points away from } v,\\
-1, & e \text{ points toward } v.
\end{cases}
\end{equation}
Eq.~\eqref{mac1} is precisely the divergenceless condition for the coarse-grained discrete flow at every internal macroscopic vertex~\cite{Cui:2018dyq}.

To study the density bound, in addition to the net flow $f_e$ we define the number of microscopic wires in a macroscopic bond $e$ that are actually occupied by the flow family:
\begin{equation}
n_e:=
\sum_{\tilde e\in B_e}|j_{\tilde e}|.
\end{equation}
Since the absolute value of the label on every microscopic wire is either $0$ or $1$, and the bundle $B_e$ contains $c_e$ elementary qubit wires,
\begin{equation}
n_e
=
\sum_{\tilde e\in B_e}|j_{\tilde e}|
\leq |B_e|
=
c_e .
\label{mac2}
\end{equation}
At the same time, the triangle inequality gives
\begin{equation}
|f_e|
=
\left|
\sum_{\tilde e\in B_e}j_{\tilde e}
\right|
\leq
\sum_{\tilde e\in B_e}|j_{\tilde e}|
=
n_e ,
\end{equation}
and therefore
\begin{equation}
|f_e|\leq n_e\leq c_e .
\label{mac2prime}
\end{equation}
Here $n_e$ records the number of trajectories that actually pass through the bond $e$, whereas $f_e$ is the net oriented flow after possible cancellations between oppositely directed contributions. Eq.~\eqref{mac2} is the density constraint in the path-configuration description, while Eq.~\eqref{mac2prime} gives the corresponding capacity constraint for the discrete flow~\cite{Cui:2018dyq}.

To summarize, the discussion above was carried out on the fully resolved microscopic string diagram $\widetilde {\mathcal{G}}$. Eq.~\eqref{mic1} shows that the certified QIF trajectories satisfy local conservation at every internal microscopic process node. Eqs.~\eqref{mac1} and~\eqref{mac2} further show that, when these microscopic lines are collected according to the macroscopic tensor vertices and bond bundles, the resulting flow satisfies the macroscopic divergenceless condition and the bond-capacity bound. The family
$
\mathcal F=\{\Gamma_r\}_{r=1}^{k}
$
therefore has the basic properties of a discrete bit-thread configuration.

Strictly speaking, however, this is a sufficient-condition statement. It does not assert that every mathematically admissible discrete bit-thread configuration can necessarily be lifted to a quantum-information-flow witness in some quantum protocol. Whether such a lift exists more generally remains an open question worthy of further investigation.

\subsection{Further remarks on the trajectory correspondence between quantum information flows and bit threads}\label{appe2}

Two further points concerning the trajectory correspondence between QIFs and bit threads deserve clarification.

First, the QIF trajectories $\Gamma_r$ analyzed above are initially defined as paths on the microscopic string diagram $\widetilde{\mathcal{G}}$. Since $\widetilde{\mathcal{G}}$ itself may be regarded as an elementary-qubit-resolution refinement of the macroscopic tensor network, and each microscopic wire has unit capacity, there is no obstruction in principle to regarding the $\Gamma_r$ themselves as discrete bit threads defined on the refined network. Indeed, it is precisely this higher resolution that allows a QIF to record how a trajectory passes through the concrete string-diagram structure inside each tensor. Such information is invisible in the usual macroscopic max-flow description.
If, however, one prefers to reserve the term ``discrete bit thread'' for trajectories on the macroscopic tensor-network skeleton
$
\mathcal{G}=(V,E,c),
$
one may further forget the internal line structure of each tensor. For this purpose, one may introduce the natural coarse-graining map
\begin{equation}
\pi:\widetilde {\mathcal{G}}\longrightarrow {\mathcal{G}} ,
\label{coa}
\end{equation}
which contracts each microscopic subdiagram $R_v$ to the corresponding macroscopic vertex $v$ and records the elementary wires in each bundle $B_e$ only at the resolution of the corresponding macroscopic bond $e$. Then, for each microscopic trajectory, define
\begin{equation}
\widetilde\Gamma_r:=\pi(\Gamma_r).
\label{til}
\end{equation}
Here $\widetilde\Gamma_r$ records only the sequence of macroscopic vertices and bonds traversed by $\Gamma_r$, without resolving which particular process nodes and wires it passes through inside a given $R_v$.
This operation clearly changes only the descriptive resolution of the trajectory. Incoming and outgoing contributions within a microscopic subdiagram cancel pairwise when it is contracted to a single macroscopic vertex, and therefore no new source or sink is introduced. Likewise, identifying the bundle $B_e$ with a macroscopic bond of capacity $c_e$ cannot increase the number of elementary wires actually occupied by the trajectories. Eqs.~\eqref{mac1} and~\eqref{mac2} are precisely the corresponding statements of this preservation. Consequently,
\begin{equation}
\widetilde{\mathcal F}
=
\{\widetilde\Gamma_r\}_{r=1}^{k}
\end{equation}
is likewise a family of macroscopic discrete bit-thread trajectories satisfying the divergenceless condition and the capacity bound.
The coarse-graining here is an optional descriptive step rather than a necessary ingredient of the trajectory correspondence. Its advantage is that it forgets the particular string-diagram expansion used inside each macroscopic tensor and thereby returns attention to the usual tensor-network skeleton.

The second subtlety is more technical and concerns compatible inheritance of paths during string-diagram rewriting. The preceding proof one chosen compatible-inheritance scheme along the rewriting sequence~\eqref{rew},
and on this basis obtained the trajectory family $\mathcal F$. A given rewriting sequence, however, does not necessarily determine a unique inheritance lineage of paths.
More specifically, under a rewrite
$
D_i\Rightarrow D_{i+1},
$
an apparent through-path $\ell_i$ in the $i$th diagrammatic representation $D_i$ may admit more than one compatible successor in $D_{i+1}$. For example, one may simultaneously have
\begin{equation}
\ell_i\,\mathcal R_i\,\ell_{i+1},
\qquad
\ell_i\,\mathcal R_i\,\ell'_{i+1},
\label{tec}
\end{equation}
where $\mathcal R_i$ denotes the compatible-inheritance relation allowed by that rewriting step. Compatible inheritance is therefore, in general, a relation rather than intrinsically a single-valued map.
Put simply, this means that, even when the initial string diagram of a definite quantum protocol is fixed, and a given rewriting sequence ultimately reduces it to $k$ decoupled bare wires, tracing those $k$ bare wires backward through the sequence need not, in principle, yield a unique possible set of QIF trajectories. A complete definition and discussion of this structure can be found in Ref.~\cite{Lin:2026hpd}.
This technical subtlety also makes clear that the trajectory nonuniqueness discussed in this work has at least two distinct levels. The GHZ-assisted teleportation and single-cell AME$(6,2)$ examples discussed above exhibit the more physical kind of nonuniqueness: the same resource state and transfer task may admit different implementing protocols, which give different entanglement-network string diagrams and hence different QIF configurations. Eq.~\eqref{tec}, by contrast, describes a finer, technical source of possible nonuniqueness associated with compatible inheritance itself. However, at least for the HaPPY constructions considered here, this latter type of nonuniqueness does not arise.



\begin{thebibliography}{99}
	\bibitem{Maldacena:1997re} J.~M.~Maldacena,
``The Large N limit of superconformal field theories and supergravity,''
Adv. Theor. Math. Phys. \textbf{2}, 231-252 (1998)
[arXiv:hep-th/9711200 [hep-th]].


	\bibitem{Gubser:1998bc} S.~S.~Gubser, I.~R.~Klebanov and A.~M.~Polyakov,
``Gauge theory correlators from noncritical string theory,''
Phys. Lett. B \textbf{428}, 105-114 (1998)
[arXiv:hep-th/9802109 [hep-th]].


	\bibitem{Witten:1998qj} E.~Witten,
``Anti-de Sitter space and holography,''
Adv. Theor. Math. Phys. \textbf{2}, 253-291 (1998)
[arXiv:hep-th/9802150 [hep-th]].




	\bibitem{Ryu:2006bv} S.~Ryu and T.~Takayanagi,
``Holographic derivation of entanglement entropy from AdS/CFT,''
Phys. Rev. Lett. \textbf{96}, 181602 (2006)
[arXiv:hep-th/0603001 [hep-th]].


	\bibitem{Ryu:2006ef} S.~Ryu and T.~Takayanagi,
``Aspects of Holographic Entanglement Entropy,''
JHEP \textbf{08}, 045 (2006)
[arXiv:hep-th/0605073 [hep-th]].


	\bibitem{Hubeny:2007xt} V.~E.~Hubeny, M.~Rangamani and T.~Takayanagi,
``A Covariant holographic entanglement entropy proposal,''
JHEP \textbf{07}, 062 (2007)
[arXiv:0705.0016 [hep-th]].




	\bibitem{Freedman:2016zud} M.~Freedman and M.~Headrick,
	``Bit threads and holographic entanglement,''
	Commun. Math. Phys. \textbf{352}, no.1, 407-438 (2017)
	[arXiv:1604.00354 [hep-th]].
	\bibitem{Cui:2018dyq} S.~X.~Cui, P.~Hayden, T.~He, M.~Headrick, B.~Stoica and M.~Walter,
	``Bit Threads and Holographic Monogamy,''
	Commun. Math. Phys. \textbf{376}, no.1, 609-648 (2019)
	[arXiv:1808.05234 [hep-th]].
	\bibitem{Headrick:2017ucz} M.~Headrick and V.~E.~Hubeny,
	``Riemannian and Lorentzian flow-cut theorems,''
	Class. Quant. Grav. \textbf{35}, no.10, 10 (2018)
	[arXiv:1710.09516 [hep-th]].
	\bibitem{Headrick:2022nbe} M.~Headrick and V.~E.~Hubeny,
``Covariant bit threads,''
JHEP \textbf{07}, 180 (2023)
[arXiv:2208.10507 [hep-th]].




\bibitem{Abramsky:2004doh}
S.~Abramsky and B.~Coecke,
``A categorical semantics of quantum protocols,''
[arXiv:quant-ph/0402130 [quant-ph]].

\bibitem{Abramsky:2008qkz}
S.~Abramsky and B.~Coecke,
``Categorical quantum mechanics,''
[arXiv:0808.1023 [quant-ph]].

\bibitem{Coecke:2005clw}
B.~Coecke,
``Kindergarten Quantum Mechanics,''
[arXiv:quant-ph/0510032 [quant-ph]].




\bibitem{Coecke:2008lcg}
B.~Coecke and R.~Duncan,
``Interacting Quantum Observables,''
Lect. Notes Comput. Sci. \textbf{5126}, 298-310 (2008)

\bibitem{Duncan:2009ocf}
R.~Duncan and B.~Coecke,
``Interacting quantum observables: categorical algebra and diagrammatics,''
New J. Phys. \textbf{13}, no.4, 043016 (2011)
[arXiv:0906.4725 [quant-ph]].

\bibitem{vandeWetering:2020giq}
J.~van de Wetering,
``ZX-calculus for the working quantum computer scientist,''
[arXiv:2012.13966 [quant-ph]].

\bibitem{Kissinger:2024pqs}
A.~Kissinger and J.~van de Wetering,
``Picturing Quantum Software: An Introduction to the ZX-Calculus and Quantum Compilation,’'
Preprint (2024).
\url{https://github.com/zxcalc/book}.


\bibitem{Coecke:2017dti}
B.~Coecke and A.~Kissinger,
``Picturing Quantum Processes,''
Cambridge University Press, 2017,
ISBN 978-1-316-21931-7




\bibitem{Coecke:2004sxv}
B.~Coecke,
``The logic of entanglement,''
[arXiv:quant-ph/0402014 [quant-ph]].

\bibitem{Coecke:2005bin}
B.~Coecke,
``Quantum information-flow, concretely, and axiomatically,''
[arXiv:quant-ph/0506132 [quant-ph]].




\bibitem{Joyal:1991esw}
A.~Joyal and R.~Street,
``The geometry of tensor calculus, I,''
Adv. Math. \textbf{88}, no.1, 55-112 (1991)



\bibitem{note:planck-thickness}
The familiar intuition that bit threads have a ``Planck-scale thickness'' already suggests that what matters need not be strict continuity itself, but rather an underlying countable and composable structure.





\bibitem{Swingle:2009bg}
B.~Swingle,
``Entanglement Renormalization and Holography,''
Phys. Rev. D \textbf{86}, 065007 (2012)
[arXiv:0905.1317 [cond-mat.str-el]].

\bibitem{Swingle:2012wq}
B.~Swingle,
``Constructing holographic spacetimes using entanglement renormalization,''
[arXiv:1209.3304 [hep-th]].

\bibitem{Pastawski:2015qua}
F.~Pastawski, B.~Yoshida, D.~Harlow and J.~Preskill,
``Holographic quantum error-correcting codes: Toy models for the bulk/boundary correspondence,''
JHEP \textbf{06}, 149 (2015)
[arXiv:1503.06237 [hep-th]].

\bibitem{Hayden:2016cfa}
P.~Hayden, S.~Nezami, X.~L.~Qi, N.~Thomas, M.~Walter and Z.~Yang,
``Holographic duality from random tensor networks,''
JHEP \textbf{11}, 009 (2016)
[arXiv:1601.01694 [hep-th]].

\bibitem{Bao:2018pvs}
N.~Bao, G.~Penington, J.~Sorce and A.~C.~Wall,
``Beyond Toy Models: Distilling Tensor Networks in Full AdS/CFT,''
JHEP \textbf{11}, 069 (2019)
[arXiv:1812.01171 [hep-th]].




\bibitem{Bennett:1992tv}
C.~H.~Bennett, G.~Brassard, C.~Crepeau, R.~Jozsa, A.~Peres and W.~K.~Wootters,
``Teleporting an unknown quantum state via dual classical and Einstein-Podolsky-Rosen channels,''
Phys. Rev. Lett. \textbf{70}, 1895-1899 (1993)

\bibitem{Nielsen:2012yss}
M.~A.~Nielsen and I.~L.~Chuang,
``Quantum Computation and Quantum Information,''
Cambridge University Press, 2012,
ISBN 978-0-521-63503-5

\bibitem{note2}
 For generic quantum states, the von Neumann entropy has its standard operational interpretation in the asymptotic limit of many copies~\cite{Nielsen:2012yss}.
Ref.~\cite{Bao:2018pvs} pointed out that, for large-$N$ holographic states with semiclassical gravitational duals, the large-$N$ limit plays an analogous concentration role in information theory, causing smooth one-shot entropies to approach the von Neumann entropy at leading order. This provides reasonable grounds for discussing, in the holographic setting, the operational meaning of the RT entropy at the level of a single copy of the state.


\bibitem{Lin:2026hpd}
Y.~Y.~Lin and C.~Y.~Li,
``Quantum Information Flow under String-Diagram Rewriting,''
[arXiv:2608.09823 [quant-ph]].


\bibitem{note3}
For a deterministic protocol, the same initial trajectory must be obtained consistently for every measurement branch.

\bibitem{Karlsson:1998opa}
A.~Karlsson and M.~Bourennane,
``Quantum teleportation using three-particle entanglement,''
Phys. Rev. A \textbf{58}, no.6, 4394 (1998)

\bibitem{Hillery:1998yq}
M.~Hillery, V.~Buzek and A.~Berthiaume,
``Quantum secret sharing,''
Phys. Rev. A \textbf{59}, 1829 (1999)
[arXiv:quant-ph/9806063 [quant-ph]].

\bibitem{Hillebrand:2011thesis}
A.~Hillebrand,
``Quantum Protocols involving Multiparticle Entanglement and their
Representations in the zx-calculus,''
MSc thesis, University of Oxford (2011).


\bibitem{note4}
At a more technical level, even for a fixed rewriting sequence, compatible inheritance itself may admit more than one valid choice of QIF trajectories~\cite{Lin:2026hpd}. 
We do not claim that these process-theoretic mechanisms exhaust all sources of nonuniqueness of mathematical maximal-flow configurations. 



\bibitem{Harper:2022sky}
J.~Harper,
``Perfect tensor hyperthreads,''
JHEP \textbf{09}, 239 (2022)
[arXiv:2205.01140 [hep-th]].

\bibitem{Bao:2023til}
N.~Bao and G.~Suer,
``Holographic entanglement distillation from the surface state correspondence,''
JHEP \textbf{01}, 091 (2024)
[arXiv:2311.07649 [hep-th]].

\bibitem{Lin:2022flo}
Y.~Y.~Lin and J.~C.~Jin,
``Thread/State correspondence: from bit threads to qubit threads,''
JHEP \textbf{02}, 245 (2023)
[arXiv:2210.08783 [hep-th]].


\bibitem{Fattal:2004frh}
D.~Fattal, T.~S.~Cubitt, Y.~Yamamoto, S.~Bravyi and I.~L.~Chuang,
``Entanglement in the stabilizer formalism,''
[arXiv:quant-ph/0406168 [quant-ph]].


\bibitem{Dehaene:2003thc}
J.~Dehaene and B.~D.~Moor,
``Clifford group, stabilizer states, and linear and quadratic operations over GF(2),''
Phys. Rev. A \textbf{68}, no.4, 042318 (2003)
[arXiv:quant-ph/0304125 [quant-ph]].


\bibitem{Aaronson:2004xuh}
S.~Aaronson and D.~Gottesman,
``Improved simulation of stabilizer circuits,''
Phys. Rev. A \textbf{70}, no.5, 052328 (2004)
[arXiv:quant-ph/0406196 [quant-ph]].


\bibitem{Nezami:2016zni}
S.~Nezami and M.~Walter,
``Multipartite Entanglement in Stabilizer Tensor Networks,''
Phys. Rev. Lett. \textbf{125}, 241602 (2020)
[arXiv:1608.02595 [quant-ph]].



\bibitem{Almheiri:2014lwa}
A.~Almheiri, X.~Dong and D.~Harlow,
``Bulk Locality and Quantum Error Correction in AdS/CFT,''
JHEP \textbf{04}, 163 (2015)
[arXiv:1411.7041 [hep-th]].

\bibitem{Dong:2016eik}
X.~Dong, D.~Harlow and A.~C.~Wall,
``Reconstruction of Bulk Operators within the Entanglement Wedge in Gauge-Gravity Duality,''
Phys. Rev. Lett. \textbf{117}, no.2, 021601 (2016)
[arXiv:1601.05416 [hep-th]].

\bibitem{Harlow:2018fse}
D.~Harlow,
``TASI Lectures on the Emergence of Bulk Physics in AdS/CFT,''
PoS \textbf{TASI2017}, 002 (2018)
[arXiv:1802.01040 [hep-th]].


\bibitem{Helwig:2013qoq}
W.~Helwig,
``Absolutely Maximally Entangled Qudit Graph States,''
[arXiv:1306.2879 [quant-ph]].



\bibitem{Munne:2022ubc}
G.~A.~Munn{\'e}, V.~Kasper and F.~Huber,
``Engineering holography with stabilizer graph codes,''
npj Quantum Inf. \textbf{10}, no.1, 51 (2024)
[arXiv:2209.08954 [quant-ph]].


\bibitem{Nest:2004khg}
M.~V.~Nest, J.~Dehaene and B.~D.~Moor,
``Graphical description of the action of local Clifford transformations on graph states,''
Phys. Rev. A \textbf{69}, no.2, 022316 (2004)
[arXiv:quant-ph/0308151 [quant-ph]].


\bibitem{Agon:2018lwq}
C.~A.~Ag\'on, J.~De Boer and J.~F.~Pedraza,
``Geometric Aspects of Holographic Bit Threads,''
JHEP \textbf{05}, 075 (2019)
[arXiv:1811.08879 [hep-th]].


\bibitem{Backens:2013hto}
M.~Backens,
``The ZX-calculus is complete for stabilizer quantum mechanics,''
New J. Phys. \textbf{16}, no.9, 093021 (2014)
[arXiv:1307.7025 [quant-ph]].

\bibitem{Backens:2015nhm}
M.~Backens,
``Making the stabilizer ZX-calculus complete for scalars,''
EPTCS \textbf{195}, 17-32 (2015)
[arXiv:1507.03854 [quant-ph]].
\bibitem{Coecke:2013tutorial}
B.~Coecke and R.~Duncan,
``Tutorial: Graphical calculus for quantum circuits,''
in \textit{Reversible Computation},
Lecture Notes in Computer Science \textbf{7581},
pp.~1--13,


\bibitem{Choi:1975nug}
M.~D.~Choi,
``Completely positive linear maps on complex matrices,''
Linear Algebra Appl. \textbf{10}, no.3, 285-290 (1975)

\bibitem{Jamiolkowski:1972pzh}
A.~Jamio{\l}kowski,
``Linear transformations which preserve trace and positive semidefiniteness of operators,''
Rept. Math. Phys. \textbf{3}, 275-278 (1972)



\bibitem{Wood:2011zvw}
C.~J.~Wood, J.~D.~Biamonte and D.~G.~Cory,
``Tensor networks and graphical calculus for open quantum systems,''
Quant. Inf. Comput. \textbf{15}, no.9-10, 0759-0811 (2015)
[arXiv:1111.6950 [quant-ph]].

\bibitem{Biamonte:2011aoz}
J.~D.~Biamonte, S.~R.~Clark and D.~Jaksch,
``Categorical Tensor Network States,''
AIP Adv. \textbf{1}, 042172 (2011)
[arXiv:1012.0531 [quant-ph]].

\bibitem{Meznaric:2012yae}
S.~Meznaric and J.~Biamonte,
``Tensor Networks for Entanglement Evolution,''
[arXiv:1204.3599 [quant-ph]].


\bibitem{Duncan:2009nrf}
R.~Duncan and S.~Perdrix,
``Graphs States and the necessity of Euler Decomposition,''
[arXiv:0902.0500 [quant-ph]].


\bibitem{Faulkner:2013ana}
T.~Faulkner, A.~Lewkowycz and J.~Maldacena,
``Quantum corrections to holographic entanglement entropy,''
JHEP \textbf{11}, 074 (2013)
[arXiv:1307.2892 [hep-th]].


\bibitem{Engelhardt:2014gca}
N.~Engelhardt and A.~C.~Wall,
``Quantum Extremal Surfaces: Holographic Entanglement Entropy beyond the Classical Regime,''
JHEP \textbf{01}, 073 (2015)
[arXiv:1408.3203 [hep-th]].



\bibitem{Agon:2021tia}
C.~A.~Ag{\'o}n and J.~F.~Pedraza,
``Quantum bit threads and holographic entanglement,''
JHEP \textbf{02}, 180 (2022)
[arXiv:2105.08063 [hep-th]].

\bibitem{Rolph:2021hgz}
A.~Rolph,
``Quantum bit threads,''
SciPost Phys. \textbf{14}, no.5, 097 (2023)
[arXiv:2105.08072 [hep-th]].

\bibitem{Headrick:2025awv}
M.~Headrick, S.~R.~Kasireddy and A.~Rolph,
``Quantum bit threads and the entropohedron,''
JHEP \textbf{04}, 196 (2026)
[arXiv:2510.22601 [hep-th]].



\bibitem{Chen:2022wvy}
L.~Chen, K.~Ji, H.~Zhang, C.~Shen, R.~Wang, X.~Zeng and L.~Y.~Hung,
``CFTD from TQFTD+1 via Holographic Tensor Network, and Precision Discretization of CFT2 ,''
Phys. Rev. X \textbf{14}, no.4, 041033 (2024)
[arXiv:2210.12127 [hep-th]].

\bibitem{Hung:2024gma}
L.~Y.~Hung and Y.~Jiang,
``Building up quantum spacetimes with BCFT Legos,''
[arXiv:2404.00877 [hep-th]].

\bibitem{Hung:2025vgs}
L.~Y.~Hung, Y.~Jiang and B.~X.~Lao,
``Universal Structures and Emergent Geometry from Large-$c$ BCFT Ensemble,''
[arXiv:2504.21660 [hep-th]].

\bibitem{Geng:2025efs}
H.~Geng, L.~Y.~Hung and Y.~Jiang,
``It from ETH: multi-interval entanglement and replica wormholes from large-c BCFT ensemble,''
JHEP \textbf{07}, 262 (2026)
[arXiv:2505.20385 [hep-th]].

\bibitem{notex}
More generally, the formalization of QIF does not, even in principle, rely on the dagger-compact categorical setting historically central to CQM. Its essential ingredients are only a process diagram with well-defined categorical semantics, semantics-preserving local rewrites, and compatible inheritance of a target simple line-like factor under those rewrites~\cite{Lin:2026hpd}. 
In fact, the categorical structure used to describe the prepared quantum structure may, in principle, even be allowed to differ from that used to describe the experimental strategy~\cite{Lin:2026hpd}.






%

\end{thebibliography}
\end{document}